\documentstyle[12pt]{article} 
\input amssym.def
\input amssym.tex

\def\goth{\frak}
\def\double{\Bbb}

\def\cc{{\double C}}
\def\nn{{\double N}}
\def\rr{{\double R}}
\def\zz{{\double Z}}

\def\mm{{{\cal M}}}

\def\ep{{\cal E}}

\def\dda{{\,\hbox{$\partial_{\!\! A}$}}}
\def\ddd{{\,\hbox{$/\!\!\!\partial$}}}
\def\ddda{{\,\hbox{$/\!\!\!\partial_{\!\! A}$}}}

\def\ot{\otimes}
\def\op{\oplus}

\def\bb{\begin{eqnarray}}
\def\ee{\end{eqnarray}}
\def\eee{\nonumber\end{eqnarray}}

 \newcommand{\gam}{\gamma}
 \newcommand{\alp}{\alpha}
 \newcommand{\bet}{\beta}

 \newcommand{\gamn}{\gamma^{\nu}}
 \newcommand{\gamm}{\gamma^{\mu}}

 \newcommand{\om}{\omega}

 \newcommand{\eps}{\varepsilon}

 \newcommand{\sig}{\sigma}

 \newcommand{\sgn}{\mathrm{sgn}}
 \newcommand{\gab}{g^{\alpha\beta}}

\newtheorem{definition}{Definition}[section]
\newtheorem{lemma}{Lemma}[section]
\newtheorem{satz}{Theorem}[section]
\newtheorem{korollar}[satz]{Corollary}
\newtheorem{proposition}{Proposition}[section]

\begin{document}

%\hsize 15truecm
%\vsize 20truecm
\font\twelve=cmbx10 at 13pt \font\eightrm=cmr8
\def\petit{\def\rm{\fam0\eightrm}}
\baselineskip 18pt

\begin{titlepage}
\title{Gauge Theories of Dirac Type}
\author {J\"urgen Tolksdorf\thanks{email: juergen.tolksdorf@mis.mpg.de}\\
Max Planck Inst. of Mathematics in the Sciences\\ Leipzig,
Germany\and
Torsten Thumst\"adter\thanks{thum@euler.math.uni-mannheim.de}\\
Inst. of Mathematics\\ University of Mannheim, Germany}
\date{November 02, 2005}
\maketitle

\begin{abstract}
A specific class of gauge theories is geometrically described in
terms of fermions. In particular, it is shown how the geometrical
frame presented naturally includes spontaneous symmetry breaking of
Yang-Mills gauge theories without making use of a Higgs
potential. In more physical terms, it is shown that the Yukawa
coupling of fermions, together with gravity, necessarily yields a
symmetry reduction provided the fermionic mass is considered as a
globally well-defined concept. The structure of this symmetry
breaking is shown to be compatible with the symmetry breaking that
is induced by the Higgs potential of the minimal Standard Model.
As a consequence, it is shown that the fermionic mass has a simple
geometrical interpretation in terms of curvature and that the
(semi-classical) ``fermionic vacuum'' determines the intrinsic
geometry of space-time. We also discuss the issue of ``fermion
doubling'' in some detail and introduce a specific projection
onto the ``physical sub-space'' that is motivated from the
Standard Model.
\end{abstract}

\vspace{0.5cm}

\begin{tabbing}
\>Spontaneous Symmetry Breaking, Fermionic Mass
 Operator
\end{tabbing}

\vspace{0.15cm}


\end{titlepage}

\section{Introduction}
The aim of this paper is to put emphasis on the role of fermions
in a geometrically unified description of different kinds of gauge
theories as, for instance, Yang-Mills and Einstein's theory.
Especially, we discuss in some detail the role of the ``Yukawa
coupling'' of fermions with respect to the mechanism of
spontaneous symmetry breaking. This may provide us with a better
geometrical understanding of the relation between inertia and
gravity.\\

Let us start out with some general remarks on the notion of
``gauge symmetry''. The notion of gauge symmetry, in general
terms, expresses certain redundancies in the mathematical description
of the interactions considered. In mathematics, by gauge
theory one usually refers to gauge theories of the Yang-Mills type
with the underlying geometry given by a principal G-bundle over
a smooth orientable (compact) manifold endowed, in addition, with
a (semi-)Riemannian structure (see, for instance, in \cite{Ble:81},
\cite{MM:92}, \cite{MMF:95}, \cite{Nab:00} and \cite{Trau:80}).
This notion of gauge theory,
however, is clearly far too restrictive when considered from a
physical point of view. For instance, gravity is also usually
regarded as a kind of gauge theory though it is certainly not of
the Yang-Mills type. The underlying geometrical structure of gravity,
regarded as a gauge theory, is that of a fiber bundle naturally
associated with the frame bundle of the base manifold $\mm$ with
typical fiber given by ${\rm GL}(n)/{\rm SO}(p,q)$. Here,
respectively, ${\rm dim}(\mm)\equiv n=p+q$ equals the dimension of
the oriented base manifold and $s=p-q$ is the signature. The
bundle structure of the two gauge theories is obviously very
different. In contrast to Yang-Mills theory, the bundle structure
of gravity is fully determined (modulo diffeomorphisms) by fixing
the (topology of the) base manifold and the signature $s$. In this
sense, the bundle structure in Einstein's theory of gravity is
more natural then in the Yang-Mills theory. Moreover, the
mathematical notion of a local trivialization has a physical
meaning in the case of gravity, however, not in Yang-Mills gauge
theories (there is no ``exponential map'' defined in Yang-Mills
theories for, in contrast to gravity, Yang-Mills connections only
determine second order vector fields but no spray fields).\\

The respective Lagrangian densities of gravity and Yang-Mills
gauge theory differ in that the former is known to be linear in
the curvature of the base manifold whereas the latter is
quadratic in the curvature of the bundle space. This difference is
known to yield far-reaching consequences, for example, when
quantization is taken into account. But also on the purely
``classical'' level (i.e. gravity and electromagnetism) there are
fundamental differences in these two kinds of gauge theories. For
example, electromagnetism (more general, Yang-Mills gauge theories
over even-dimensional base manifolds) is known to be scale
invariant but not invariant with respect to the action of the
diffeomorphism group (except isometries). In contrast, gravity is
covariant with respect to diffeomorphisms but not scale invariant.
Of course, despite these profound mathematical and physical
differences there are, nonetheless, formal similarities between
these two types of gauge theories. Especially, the dynamics that
is defined by both theories can be expressed with respect to
top-forms on the base manifold with the property of being
invariant with respect to the action of their respective symmetry
groups. A natural question then is whether these two fundamental
kinds of gauge theories have a common geometrical root.\\

Of course, over the last decades there have been various
attempts to geometrically unify gravity with Yang-Mills
gauge theory. This holds true for string theory and, in
particular, for various aspects of non-commutative
geometry, see, for example, \cite{CFG:95}, \cite{CC:97},
\cite{Con:96}, \cite{Oku:00} and the corresponding references
there. The fruitful idea to consider the Higgs boson of the
Standard Model as an integral part of the Yang-Mills theory goes
back to fundamental works as, for example, \cite{Con:88},
\cite{CL:90}, \cite{Coq:89} and \cite{CEV:'91}. It is well-known
that this idea actually has had a tremendous impact on a vast
variety of papers of the same theme (see, for instance, \cite{FGLV:97},
\cite{KMO:98}, \cite{MO:94} and \cite{MO:96} in the context of
non-commutative geometry, or \cite{HPS:91} and \cite{NS:91} in the
case of ``super-algebras''). Basically, of all
of these geometrical descriptions of gauge theory use the purely
algebraic content of gauge theories of the Yang-Mills type (e.g.: the
exterior differential is an nilpotent derivation and a connection
is the sum of the latter and a one-form) as there starting point.
However, gravity seems not to fit in this basic algebraic sight.
Also, spontaneous symmetry breaking is described only in terms of
(the algebraic aspects of) Yang-Mills gauge theories without using
gravity. The notion of fermions only arises because in the
algebraic context the exterior differential is defined in
terms of specific generalizations of the notion of a Dirac
operator. These purely algebraic generalizations of the latter,
however, seem to have no geometrical counter part (see, for
instance, the ``internal Dirac operator'' in the geometrical
description of the Standard Model in terms of
``almost commutative models'', \cite{IS:95}).\\

In the following we shall discuss a specific class of gauge
theories including Einstein's theory of gravity and (spontaneously
broken) Yang-Mills theory from the point of view of fermions. The
latter will be geometrically treated as certain Hermitian vector
bundles over arbitrary smooth orientable manifolds of even
dimension. These ``fermion bundles'' correspond to a global
specification of a certain class of first order differential
operators, called ``Dirac type operators''. We introduce a
canonical mapping which associates with every Dirac type operator
a specific top-form on the base manifold. This canonical mapping
is then referred to as the ``Dirac-Lagrangian''
on the setup to be discussed. The Dirac-Lagrangian turns out to be
equivariant with respect to bundle equivalence. In particular, it
is invariant with respect to the action of the Yang-Mills and the
Einstein-Hilbert gauge group. The diffeomorphism group of the base
manifold is naturally included by the pull-back action. We also
consider a distinguished class of Dirac type operators within
this setup. The corresponding top-form associated with these Dirac
type operators is shown to define a spontaneously broken gauge
theory without referring to a Higgs potential. In more physical
terms, it is shown that the ``Yukawa coupling'' together with
gravity yields a symmetry reduction which is compatible with the
symmetry breaking induced by a Higgs potential of the form used in
the (minimal) Standard Model of Particle Physics. In fact, the
latter is shown to be naturally generated by a ``fluctuation of
the fermionic vacuum''. We will also reformulate the notion of
``unitary gauge'' in terms of Dirac type operators and give
necessary and sufficient conditions for its global existence.\\

The geometrical description of gauge theories discussed in the
present paper is a considerable refinement of the geometrical
frame that has been introduced in \cite{Tol:98} in the case of
elliptic Dirac type operators on a smooth even-dimensional closed
Riemannian Spin-manifold. In contrast to the latter we will consider
in this paper the more physically appropriate case of arbitrary signature
and non-compact manifolds. Also, we do not assume that ``space-time''
has a spin structure (please, see below). For this, however,
we will focus on (globally defined) densities instead of action
functionals. Accordingly, we have to demand that the densities
themselves are covariant with respect to the underlying symmetry
action and thus well-defined on the appropriate moduli spaces.
This is achieved mainly since the densities in question are derived
from evaluating a natural object (within the frame considered)
with respect to specific first order differential operators. As a
consequence, one ends up with densities which are linear in the
curvature of the base manifold and quadratic in the curvature of
the bundle space. For instance, it is shown that the total
curvature of the ``fermionic vacuum'' decomposes into the sum of
the curvature of the base manifold together with the (square of
the) fermionic mass operator. Also a basic difference relative to
the frame considered in loc. sit. (and subsequent papers thereof)
is that all bundles, including the Higgs and the Yang-Mills bundle,
are considered as specific sub-bundles of the fermion bundle (resp.
of the bundle of endomorphisms of the latter). The fermion density
will be considered as a specific mapping on the affine set of all
Dirac type operators on a fermion bundle. Here, we also discuss the
issue of the doubling of the fermionic degrees of freedom that is
necessary to apply the general Bochner-Lichnerowizc-Weizenb\"ock
formula.\\

Finally, we want to comment on the notion of ``fermions'' without
assuming the existence of spin structures. At least in the so-called
``semi-classical approximation'' of a full quantum field theory it is
common to geometrically treat the ``states of a fermion'' as sections
of a (twisted) spinor bundle over space-time. For this, of course, the
topology of space-time must guarantee the existence of a spin
structure (i.e. the vanishing of the second Stiefel-Whitney classes).
Moreover, together with the assumption of global hyperbolicity the
existence of spin structures in four dimensions is known to be
equivalent to the triviality of the frame bundle of space-time
(``Geroch's Theorem''). Therefore, the existence of a spin
structure provides severe restrictions to the topology of
space-time. However, the experiments performed to demonstrate that
the double cover of the (proper orthochroneous) Lorentz group
is more fundamental are purely local in nature. Also, in order to
obtain a topologically non-trivial statement about the existence
of spin structures, space-time has to be covered by at least three
(trivializing) local charts. This, of course, rises the question
of the physical sense of ``locality'' in this context to give
the mathematical construction a physical meaning. Hence, from our
point of view, the assumption of the existence of a spin structure
is a purely mathematical one without a physically meaningful
counterpart. In fact, in this respect the notion of ``locality'',
as it is used in mathematics, seems physically as spurious as in
the case of Yang-Mills gauge theories which do not provide any
scale. Basically, this is the reason to consider in this work the
more general notion of ``Clifford module bundles'' instead of
``twisted spinor bundles'' as an appropriate geometrical background.
In contrast to the latter, the existence of Clifford module bundles
yields no more topological restrictions on space-time than the
existence of a metric itself. For instance, the bundle of
Grassmann algebras severs as a natural Clifford module bundle
for every space-time manifold. However, the topology of the
Clifford module bundles cannot be arbitrary. The physical
interpretation of the sections of Clifford module bundles in
terms of the states of fermions yields restrictions to the
topology of the considered Clifford module bundles
(please, see below).\\

The paper is organized as follows: In the next section we
introduce the concept of fermion bundles as a specific class of
Clifford module bundles and define Dirac type
gauge theories. In the third section we consider a distinguished
class of such gauge theories and discuss spontaneous symmetry
breaking in this context. In the fourth section we introduce the
fermionic density within the presented geometrical setup and
discuss the issue of fermionic doubling. In the fifth section we
want to specify what we mean by a ``fluctuation of the
fermionic vacuum''. This is done in terms of yet another class of
Dirac type operators. Finally, in section six we close with an
outlook. In an appendix we present a detailed
proof of the explicit form of ``simple type Dirac operators'' of
arbitrary signature, for these operators turn out to be
fundamental, e.g., in our discussion of spontaneous symmetry
breaking.

%%%%%%%%%%%%%%%%%%%%%%%%%%%%%%%%%%%%%%%%%%%%%%%%%%%%%%%%%%%%%%%%%%%%%%%%%%%%%%%%

\section{Fermion Bundles and Dirac Type Gauge Theories}
In this section we introduce a specific class of Clifford module
bundles which will serve as our geometrical background for gauge
theories. On this background there exists a canonical mapping
which permits to associate with the local data of a fermion bundle
a specific top form on the base manifold. This top form turns out
to be equivariant with respect to the automorphism group of the
underlying geometrical structure.

\subsection{Fermion Bundles, Dirac Type Operators and Connections}
In this sub-section we define our notion of {\it fermion bundles}
as a specific class of Clifford module bundles.
For this let $\xi := (\ep,\pi_\ep,\mm)$ be a smooth
complex vector bundle with total space $\ep,$ base manifold $\mm$
and projection map $\pi_\ep:\,\ep\rightarrow\mm.$ The rank, ${\rm
rk}(\xi)\in\nn,$ of the bundle is ${\rm N}\geq 1.$ In what follows
the base manifold is assumed to be  orientable and of even
dimension $n\equiv 2k.$ As a topological space $\mm$ is a
para-compact and (simply-) connected Hausdorff-space. On this
geometrical background we consider the following local data:
\bb
\label{dopsdata}
({\rm G},\rho_{\rm F},D).
\ee
Here, G is a semi-simple, compact and real Lie group and
$\rho_{\rm F}:\,{\rm G}\rightarrow{\rm SU}({\rm N}_{\rm F})$ is a
unitary and faithful representation thereof. Moreover,
$D:\,\Gamma(\xi)\rightarrow\Gamma(\xi)$ is a first order
differential operator, acting on sections of the bundle $\xi$ such
that the bilinear extension $g_{\rm M}$ of the mapping
$(df,dh)\mapsto{\rm tr}([D,f][D,h])/{\rm rk}(\xi)$ is
non-degenerated for all smooth functions
$f,g\in{\cal C}^\infty(\mm)$. The operator $D$ is said to have
the signature $s\in\zz,$ provided that the quadratic form associated
with the (semi-)Riemannian metric $g_{\rm M}$ has signature $s.$
The mapping $g_{\rm M}$ corresponds to a
section of the ``Einstein-Hilbert bundle'' $\xi_{\rm EH}:=
(F\mm\times_{\rm GL(n)}{\rm GL}(n)/{\rm O}(p,q),\pi_{\rm EH},\mm),$
with, respectively, $F\mm$ the total space of the frame bundle
${\cal FM}$ of the base manifold $\mm$ and
$n\equiv p+q,\, s\equiv p-q.$\\

Let $\tau_{\rm Cl}$ be the algebra bundle of Clifford algebras
which are point-wise generated by $(\tau^*_{\rm M},g_{\rm M}),$
with $\tau^*_{\rm M}$ being the cotangent bundle of $\mm.$ By the
very definition, the principal symbol of the operator $D$ induces
a Clifford (left) action $\gamma:\tau_{\rm Cl}\rightarrow{\rm End}(\xi)$
via the mapping
\bb
\tau_{\rm Cl}\times\xi&\longrightarrow&\xi\cr
(df,\goth{z})&\mapsto&[D,f]\goth{z},
\ee
for all smooth functions
$f\in{\cal C}^\infty(\mm).$ As a consequence, the algebra bundle
of endomorphisms on $\xi$ globally decomposes as \bb
\label{endrelation} {\rm End}(\xi) \simeq \tau_{\rm
Cl}^\cc\ot_\mm{\rm End}_{\rm Cl}(\xi). \ee Here, ${\rm End}_{\rm
Cl}(\xi)\subset{\rm End}(\xi)$ denotes the sub-bundle of
endomorphisms which super-commute with the Clifford action
$\gamma$ (c.f., for instance, in \cite{ABS:64} and \cite{BGV:96}).

\begin{definition}
The vector bundle $\xi\equiv\xi_{\rm F}$ is called a ``fermion
bundle'' with respect to the (local) data (\ref{dopsdata}) if the
structure group of $\xi$ can be reduced to
${\rm Spin}(p,q)\times\rho_{\rm F}(\rm G).$ A
fermion bundle is called ``chiral'' provided
$\xi_{\rm F}=\xi_{\rm F}^+\op\xi_{\rm F}^-$ is $\zz_2-$graded
with respect to some involution
$\Gamma\equiv\gamma_{\rm M}\ot\chi\in\Gamma({\rm End}(\xi_{\rm F})).$
Here, the grading involution $\gamma_{\rm M}\in\Gamma(\tau_{\rm Cl}^\cc)$
is defined in terms of the (semi-)Riemannian volume form
$\mu_{\rm M}\in\Omega^n(\mm)$ that is induced by $g_{\rm M}.$
That is, $\gamma_{\rm M}\sim\gamma(\mu_{\rm M}).$
Moreover, $\xi_{\rm F}$  is called ``real'' if
all of its odd Chern classes vanish. With respect to $\Gamma$ the operator
$D$ is supposed to be odd and the representation $\rho_{\rm F}$ is assumed
to be even. In this case, $D$ is called a ``Dirac type operator'' and
(\ref{dopsdata}) a ``Dirac triple''.
\end{definition}

A fermion bundle encodes the global data of a Dirac type gauge
theory. With respect to these data we consider the set ${\cal
D}(\xi_{\rm F})$ of all Dirac type operators $D'\in{\cal
D}(\xi_{\rm F})$ satisfying the condition $[D'-D,f]\equiv 0$ for
all $f\in{\cal C}^\infty(\mm).$ The set ${\cal D}(\xi_{\rm F})$
naturally becomes an affine space with vector space $\Gamma({\rm
End}^-(\xi_{\rm F})).$ In what follows we summarize
the basic features of this affine space.\\

The affine space ${\cal A}(\xi_{\rm F})$ of linear connections on
$\xi_{\rm F}$ has a distinguished affine sub-space ${\cal A}_{\rm
Cl}(\xi_{\rm F})\subset{\cal A}(\xi_{\rm F})$ that is defined by
all linear connections which are compatible with the Clifford
action $\gamma.$ That is, $A\in{\cal A}_{\rm Cl}(\xi_{\rm F})$
defines a covariant derivative $\dda$ satisfying
$[\dda,\gamma(a)]=\gamma(\nabla^{\rm Cl}a)$ for all sections
$a\in\Gamma(\tau_{\rm Cl}^\cc)$ and $\nabla^{\rm Cl}$ being the
covariant derivative with respect to the lifted Levi-Civita
connection of $g_{\rm M}.$ Accordingly, such a connection is
referred to as a ``Clifford connection''. Hence, every $D'\in{\cal
D}(\xi_{\rm F})$ may be written as $D'=\ddda + \Phi$ where,
respectively, $\ddda\equiv\gamma\circ\dda$ is the analogue of a
twisted Spin-Dirac operator in the case where $\mm$ denotes a
spin-manifold and $\Phi\equiv D-\ddda\in\Gamma({\rm End}(\xi_{\rm
F})).$ Notice, however, that in general the zero order operator
$\Phi$ also depends on the Clifford connection $A.$ Moreover, the
relation between the two affine spaces ${\cal D}(\xi_{\rm F})$ and
${\cal A}(\xi_{\rm F})$ on a fermion bundle is given by the
(signature independent) bijection (c.f. \cite{Tol:98})
\bb
\label{dopsrelcon}
{\cal D}(\xi_{\rm F})\simeq{\cal A}(\xi_{\rm F})/{\rm ker}(\gamma).
\ee
Therefore, to each Dirac type operator
on $\xi_{\rm F}$ there corresponds an equivalence class of
connections. However, each connection class has a natural
representative that is constructed as follows: Firstly, on every
chiral fermion bundle there is a canonical odd one-form
$\Theta\in\Omega^1(\mm,{\rm End}^-(\ep))$ that is given by the
(normalized) lifted soldering form of ${\cal FM}.$ More precisely,
let $\vartheta\in\Omega_{\rm eq,hor}^1(F\mm,\rr^n)$ be the
soldering form on the (total space of the) frame bundle of $\mm.$
Here, the canonical identification $\Omega_{\rm eq,hor}^*(F\mm,\rr^n)
\simeq\Omega^*(\mm,T\mm)$ and the injection\footnote{Here,
$\Omega_{\rm eq,hor}^*(FM,\rr^n)$ denotes the ``right-equivariant'' and
``horizontal'' forms on the total space of the frame bundle of $\mm.$}
\bb
\Gamma(\tau^*_{\rm M}\ot_\mm\tau_{\rm M})\stackrel{{\rm
id}\ot\phantom{x}^\flat} {\longrightarrow}\Gamma(\tau^*_{\rm
M}\ot_\mm\tau^*_{\rm M})\hookrightarrow \Gamma(\tau^*_{\rm
M}\ot_\mm\tau_{\rm Cl})\stackrel{{\rm id}\ot\gamma}
{\longrightarrow}\Gamma(\tau^*_{\rm M}\ot_\mm{\rm End}(\xi_{\rm F}))
\ee
yields $\Theta:=\pm{\tilde\vartheta}/n$ with
${\tilde\vartheta}\equiv
\gamma\circ\vartheta^\flat\in\Omega^1(\mm,{\rm End}(\ep)).$ If
$(X_1,\ldots,X_n)$ denotes a local frame on $\mm$ and
$(X^1,\ldots,X^n)$ its dual, then\footnote{Throughout the paper
Einstein's summation convention is used in local formulas except
where this may lead to confusions.}
${\tilde\vartheta}\stackrel{loc.}{=}
X^k\ot\gamma(X_k^\flat)\ot{\rm id},$ with the usual ``musical''
isomorphism $u^\flat(v):=g_{\rm M}(u,v)$ for all $u,v\in T\mm.$ The
normalized soldering form $\Theta$ has the two basic properties:
It is covariantly constant with respect to every Clifford
connection and it induces a canonical right inverse of the
Clifford action, i.e. $\gamma\circ\,{\rm ext}_\Theta={\rm id}.$
Here, ${\rm ext}_\Theta\in
{\rm End}(\tau^*_{\mbox{\tiny$\Lambda $M}}\ot_\mm{\rm End}(\xi))$
denotes the operator of (point-wise) left-multiplication by $\Theta,$
and $\tau^*_{\mbox{\tiny$\Lambda $M}}$ is the bundle of Grassmann
algebras that is, again, generated by $\tau^*_{\mbox{\tiny M}}$.
Note that the linear equivalence
$\tau^*_{\mbox{\tiny Cl}}\simeq\tau^*_{\mbox{\tiny$\Lambda $M}}$
is used but not explicitly indicated.
Secondly, to each Dirac type operator $D'\in{\cal D}(\xi_{\rm F})$
there exists a correspondingly unique connection ${\hat A}'_{\rm D}\in{\cal
A}(\xi_{\rm F})$ such that $D'^2-\triangle'_{\mbox{\tiny
D}}\in\Gamma({\rm End}(\xi_{\rm F})).$ The second order operator
$\triangle'_{\mbox{\tiny D}}:= -{\rm
tr}({\hat\nabla}^{{T^*\!\mm}\ot\ep}\circ{\hat\nabla}^\ep)$ is called the
``Bochner-Laplacian'' of $D'$ (c.f., for example \cite{BGV:96},
\cite{BG:90}, or \cite{Gil:95}). Here, ${\hat\nabla}^\ep$ denotes
the covariant derivative that corresponds to the connection ${\hat
A}'_{\rm D}.$ As a consequence, the covariant derivative that is
defined by
\bb \label{diracconnection}
\partial_{\! D'}:={\hat\nabla}^\ep +
\Theta\wedge(D'-\gamma\circ{\hat\nabla}^\ep)
\ee
yields a connection $A'_{\mbox{\tiny D}}\in{\cal A}(\xi_{\rm F})$
which clearly represents the Dirac type operator $D',$ i.e.
$D'=\gamma\circ\partial_{\!D'}.$ We call, respectively,
$A'_{\mbox{\tiny D}}$ the Dirac connection associated with $D'$
and the one-form
\bb
\label{diracform}
\varpi'_{\!\mbox{\tiny
D}}:=\Theta\wedge(D'-\gamma\circ{\hat\nabla}^\ep)\in
\Omega^1(\mm,{\rm End}(\ep))
\ee
the ``Dirac form'' associated
with $D'\in{\cal D}(\xi_{\rm F}).$ Of course, if the connections
${\hat A}'_{\rm D}$ and $A'_{\mbox{\tiny D}}$ are identified with
the respective connection forms ${\hat\omega},
\omega'_{\mbox{\tiny D}}\in\Omega^1(\ep,T\ep),$ then
\bb
\pi_\ep^*\varpi'_{\!\mbox{\tiny D}} = \omega'_{\mbox{\tiny D}} -
{\hat\omega}.
\ee
\phantom{Here is a line}

\vspace{0.5cm}

\noindent
{\bf Remark:}\\
{\small\sl As a first order differential operator each Dirac type operator
$D$ is known to be of the (local) form: $D=\gamma^\mu(\partial_\mu
+ \omega_\mu)$ with the appropriate ``$\gamma-$matrices''
$\gamma^\mu\in{\rm End}(\cc^{2^{\rm k}})$ satisfying either of
the Clifford relations $\gamma^\mu\gamma^\nu +
\gamma^\nu\gamma^\mu \equiv \pm 2g^{\mu\nu}{\bf 1},$
and}
\bb \label{localdops}
\omega_\mu\equiv\omega^{\mbox{\tiny Cl}}_\mu\ot{\bf 1} + {\bf 1}\ot A_\mu
\pm{\mbox{\small$\frac{1}{n}$}}\,g_{\mu\nu}\sum_{0\leq k\leq n}\;
\sum_{1\leq i_1<i_2\cdots<i_k\leq n}
\gamma^\nu\gamma^{i_1}\gamma^{i_2}\cdots\gamma^{i_k}\ot\theta_{i_1 i_2\cdots i_k}.
\ee

{\small\sl Here, respectively, $\omega^{\mbox{\tiny Cl}}_\mu$
is the component of the lifted Levi-Civita form with
respect to the appropriate metric coefficients $g_{\mu\nu},$ and
$A_\mu, \theta_{i_1 i_2\cdots i_k}$ are the components of locally
defined differential forms of various degrees which take their
values in $\rho'_{\rm F}({\rm Lie(G)})\subset{\rm End}(\cc^{{\rm
N}_{\rm F}}).$ Obviously, these forms determine each specific
Dirac type operator $D\in{\cal D}(\xi_{\rm F})$ locally. More
precisely, let $\{(U_\alpha,\chi_\alpha)\,|\,\alpha\in\Lambda\}$
be a family of local trivializations of the underlying vector
bundle $\xi,$ i.e.
$\chi_\alpha:\,\pi_\ep^{-1}(U_\alpha)\stackrel{\simeq}{\rightarrow}
U_\alpha\times\cc^{\rm N}.$ Accordingly, let $\chi_{\alpha\beta}:\,
U_\alpha\cap U_\beta\rightarrow{\rm GL}({\rm N},\cc)$
be the appropriate transition functions. Then, a family of
first order differential operators $D_\alpha$ of the form
$D_\alpha=\gamma_\alpha\circ\nabla_\alpha,$ with
$\nabla_\alpha\equiv d + \omega_\alpha$ and $\omega_\alpha$
defined by (\ref{localdops}), gives rise to a Dirac type operator
$D$ on $\xi$ provided the principal symbols $\gamma_\alpha$ define
a family of Clifford mappings $\rr^{p,q}\rightarrow{\rm
End}(\cc^{{\rm 2}^{\rm k}})\simeq\cc\ot{\rm Cl}_{p,q}$ and
the transition functions take their values in the
subgroup ${\rm Spin}(p,q)\times\rho_{\rm F}({\rm G})$ such that the
family $\{(U_\alpha,D_\alpha)\,|\,\alpha\in\Lambda\}$ fulfills the
compatibility condition $D_\alpha = \chi_{\alpha\beta}\circ
D_\beta\circ\chi_{\alpha\beta}^{-1}$ for all $x\in U_\alpha\cap
U_\beta\subset\mm.$ Hence, the notion of a Dirac triple on $\xi$
(i.e. $\xi_{\rm F}$) globalizes what is encoded in the local data
specifying $D.$ In other words, the notion of a fermion bundle
simply permits globalization of the local data
$(U_\alpha,D_\alpha)_{\alpha\in\Lambda}$ usually encountered in
physics.}

\vspace{0.5cm}

\subsection{Gauge Theories of Dirac Type and their Gauge Groups}
In this sub-section we show that the geometrical setup of fermion
bundles permits to naturally introduce a specific class of gauge
theories which we call {\it gauge theories of Dirac type} (GTDT).
The corresponding gauge group is the automorphism group of the
underlying geometrical structure. It is shown that this group
decomposes into certain subgroups which can be identified with
the usual Yang-Mills gauge group, the Einstein-Hilbert gauge group
and the diffeomorphism group of the base manifold.

\begin{definition}
Two fermion bundles $\xi_{\rm F}$ and $\xi_{\rm F}'$ are
considered to be equivalent if ${\rm G}\simeq{\rm G}'$ and
$\rho_{\rm F}$ is similar to $\rho_{\rm F}'.$ Moreover, there is a
bundle isomorphism $(\alpha,\beta):\,\xi\rightarrow\xi'$ (i.e.
diffeomorphisms $\alpha:\,\mm\rightarrow\mm'$ and
$\beta:\,\ep\rightarrow\ep',$ with $\beta$ being fiber-wise
linear and $\alpha\circ\pi_\ep = \pi_{\ep'}\circ\beta$)
such that $D'=\beta\circ D\circ\beta^{-1}.$
\end{definition}

Notice that the condition $D'=\beta\circ D\circ\beta^{-1}$
actually is equivalent to $g_{\rm M'}={\alpha^{-1}}^*g_{\rm M}.$\\

The presented geometrical setup permits the formulation of a
class of gauge theories which are based on a ``universal Lagrangian''
that is covariant with respect to the action of the automorphism group
\bb
\label{totalgaugegroup}
{\cal G}_{\rm F}\equiv{\rm Aut}(\xi_{\rm F}):=
\{(\alpha,\beta)\in{\rm Diff}(\mm)\times{\rm Aut}(\ep)\,|\,
\pi_\ep\circ\beta=\alpha\circ\pi_\ep\}
\ee
of the fermion bundle in question. This group may be identified with
the group of right-equivariant automorphisms of the frame bundle
associated with $\xi_{\rm F}.$ That is,
\bb
{\cal G}_{\rm F}\simeq{\rm Aut}_{\rm eq}({\cal FE}):=
\{f\in{\rm Aut}(F\ep)\,|\,R_g\circ f=f\circ R_g,\;g\in {\rm G}_{\rm F}\}
\ee
where, respectively, ${\cal FE}\equiv(F\ep,\mm,\pi,{\rm G}_{\rm F})$ is
the associated frame bundle of the fermion bundle considered,
${\rm G}_{\rm F}\equiv{\rm Spin}(p,q)\times\rho_{\rm F}({\rm G})$ its
structure group and $R$ the right action of the latter on the
total space $F\ep$ of the frame bundle.\\

Therefore, the automorphism group (\ref{totalgaugegroup}) has several
important sub-groups. In particular, it contains the ``inner gauge group''
of the fermion bundle $\xi_{\rm F}:$
\bb
\label{innergaugegroup}
{\cal G}_{\rm in} :=
\{(\alpha,\beta)\in{\cal G}_{\rm F}\,|\, \alpha:={\rm id}_\mm\},
\ee
which may be identified with the gauge group of
${\cal FE}.$ The latter contains two mutually commuting normal
sub-groups ${\cal G}_{\rm EH}$ and ${\cal G}_{\rm YM},$ such that
${\cal G}_{\rm EH}\cap{\cal G}_{\rm YM}=\{e\}.$ Therefore,
\bb
\label{innergaugegroup}
{\cal G}_{\rm in}\simeq{\cal G}_{\rm EH}\times_\mm{\cal G}_{\rm YM}.
\ee

Here, the ``Yang-Mills gauge group'' ${\cal G}_{\rm YM}$ can be
identified with the sub-group
$\{(\alpha,\beta)\in{\rm Aut}(\xi_{\rm F})\,|\,\alpha={\rm id}_\mm,
\beta\in{\rm Aut}_{\rm Cl}(\ep)\}$ of the inner gauge group
(\ref{innergaugegroup}). Note that the Yang-Mills gauge group is
in fact an invariant sub-group of the inner gauge group. Hence,
with respect to the foregoing mentioned identification the
``Einstein-Hilbert gauge group'' ${\cal G}_{\rm EH}$ may be
identified with the quotient group ${\cal G}_{\rm in}/{\cal G}_{\rm YM}$
according to the decomposition (\ref{endrelation}).\\

Moreover, the diffeomorphism group of the base manifold $\mm$ has
a natural non-trivial embedding into ${\rm Aut}(\xi_{\rm F}).$ Indeed,
if $\xi_{\rm F}$ is merely considered as a vector bundle then one gets the
(trivial) embedding
\bb
\label{diffembedding}
{\rm Diff}(\mm)&\hookrightarrow&{\rm Aut}(\xi_{\rm F})\cr
\alpha &\mapsto&(\alpha,\beta:=\pi_\ep^*\alpha\times{\rm id}_\ep).
\ee
This embedding may actually be identified with the inclusion
according to the definition (\ref{totalgaugegroup}) of the automorphism
group and the identification
\bb
{\alpha^{-1}}^*\ep\equiv
\{(y,{\goth z})\in\mm\times\ep\,|\,\pi_\ep({\goth z})=\alpha^{-1}(y)\}=
\{({\pi^*\!\alpha}({\goth z}),{\goth z})\,|\,{\goth z}\in\ep\}\simeq\ep.
\ee

Hence, one has ${\alpha^{-1}}^*\xi_{\rm F}=({\alpha^{-1}}^*\ep,\mm,{\rm pr}_1)=
(\ep,\mm,\pi^*\alpha)$ which permits to replace
$\beta=\pi^*\alpha\times{\rm id}_\ep$ (with inverse given by ${\rm pr}_2$)
simply by $\beta:={\rm id}_\ep.$ However, since $\xi_{\rm F}$ is
a Clifford module bundle over $(\mm,g_{\rm M}),$ the embedding of
${\rm Diff}(\mm)$ into ${\rm Aut}(\xi_{\rm F})$ becomes non-trivial. In
other words, there is an inner automorphism on ${\rm End}(\ep),$ induced
by $\alpha,$ such that $\gamma'=
{\tilde\alpha}\circ\gamma\circ{\tilde\alpha}^{-1}.$
Here, $\gamma'|_{TM}\equiv\gamma|_{TM}\circ T\alpha^{-1}$ is the
Clifford action on
${\alpha^{-1}}^*\xi_{\rm F}$ that is defined with respect to
${\alpha^{-1}}^*\!g_{\rm M}$ and ${\tilde\alpha}\in{\rm End}(\ep)$ an
appropriate lift of $\alpha.$\\

As a consequence, one obtains
\bb
{\rm Diff}(\mm)&\hookrightarrow&{\rm Aut}(\xi_{\rm F})\cr
\alpha &\mapsto& (\alpha,\beta:={\tilde\alpha}).
\ee
We call the image of this embedding the ``outer gauge group'' of
the fermion bundle $\xi_{\rm F}.$ It is denoted by ${\cal G}_{\rm ex}.$\\

Finally, since ${\cal G}_{\rm in}\subset{\cal G}_{\rm F}$
is normal and ${\cal G}_{\rm in}\cap{\cal G}_{\rm ex}=\{e\},$
one ends up with the semi-direct decomposition of the automorphism
group into the gauge and diffeomorphism group, i.e.
\bb
{\cal G}_{\rm F} = {\cal G}_{\rm in}\rtimes{\cal G}_{\rm ex}.
\ee
In fact, each $g\in{\cal G}_{\rm F}$ may be written as
$g=g_{\rm in}g_{\rm ex}\in{\cal G}_{\rm in}\rtimes{\cal G}_{\rm ex}$
such that
\bb
{\cal G}_{\rm F}\ni gg'=
(g_{\rm in}g_{\rm ex})(g'_{\rm in}g'_{\rm ex})\equiv
(g_{\rm in}g_{\rm ex}g'_{\rm in}g^{-1}_{\rm ex})(g_{\rm ex}g'_{\rm ex})
\in{\cal G}_{\rm in}\rtimes{\cal G}_{\rm ex}.
\ee

We call the automorphism group ${\cal G}_{\rm F}\equiv{\rm Aut}(\xi_{\rm F})$
the ``(fermionic) gauge group'' of the fermion bundle $\xi_{\rm F}.$\\

In order to define a ${\cal G}_{\rm F}-$covariant theory (by which
we mean that symbolically ${\cal L}\circ(\alpha,\beta)=
{\alpha^{-1}}^*{\cal L}$ where ${\cal L}$ is an
appropriate ``Lagrangian density'' defining the theory) we first
consider, for a given fermion bundle $\xi_{\rm F},$ the canonical
mapping
\bb
\label{diracpot}
V_{\rm D}:\,{\cal D}(\xi_{\rm F})&\longrightarrow&{\cal C}^\infty(\mm)\cr
D'&\mapsto& {\rm tr}(D'^2 - \triangle'_{\mbox{\tiny D}})
\ee
which is called the ``Dirac potential'' on $\xi_{\rm F}$. Here again,
the second order differential operator $\triangle'_{\mbox{\tiny D}}$
denotes the Bochner-Laplacian that is uniquely defined with respect
to $D'$ such that $\triangle'_{\mbox{\tiny D}} +
(D'^2 - \triangle'_{\mbox{\tiny D}})$
is the (general) Lichnerowicz decomposition of $D'^2$
(c.f. in \cite{BGV:96}, \cite{Gil:95}).\\

The universal top form
\bb
\label{diraclagrangian}
{\cal L}_{\rm D}:\,{\cal D}(\xi_{\rm F})&\longrightarrow&\Omega^n(\mm)\cr
D'&\mapsto& \ast V_{\rm D}(D')
\ee
is called the ``Dirac-Lagrangian'' on the fermion bundle $\xi_{\rm F}.$
This canonical mapping is universal in the sense that it is indeed
covariant with respect to the action of ${\cal G}_{\rm F}$. In
particular, it is invariant with respect to the action of the
inner gauge group ${\cal G}_{\rm in}\subset{\cal G}_{\rm F}.$\\

\begin{definition}
Let $\xi_{\rm F}$ be the fermion bundle with respect to the data
$({\rm G},\rho_{\rm F},D).$ We call (the ``bosonic part'' of) the
theory which is defined by the corresponding Lagrangian density
${\cal L}_{\rm D}(D)\in\Omega^n(\mm)$ a ``gauge theory of Dirac
type''.
\end{definition}

Let again ${\cal A}(\xi_{\rm F})$ be the set of all linear
connections on $\xi_{\rm F} $ and ${\cal A}_{\rm D}(\xi_{\rm
F})\subset{\cal A}(\xi_{\rm F})$ be the subset of all connections
which yield $D$ (i.e. $\gamma\circ\nabla^\ep=D,$ with $\nabla^\ep$
a corresponding covariant derivative). Then, the top form ${\cal
L}_{\rm D}(D)\in\Omega^n(\mm)$ is indeed well-defined on the
moduli-space ${\goth M}_{\rm D}(\xi_{\rm F})\equiv{\cal A}_{\rm D}
(\xi_{\rm F})/{\cal G}_{\rm in}.$ Moreover, it transforms covariantly
with respect to the (left) action of the fermionic gauge group
${\cal G}_{\rm F},$ i.e.
\bb
{\cal L}'_{\rm D}(\beta\circ D\circ\beta^{-1}) =
({\alpha^{-1}}^*{\cal L}_{\rm D})(D).
\ee

To obtain an explicit formula for the top form ${\cal L}_{\rm D}(D)$
associated with a Dirac type operator $D,$ one could use
the generalized Bochner-Lichnerowicz-Weizenb{\"o}ck formula of
$D^2 - \triangle_{\mbox{\tiny D}}\in\Gamma({\rm End}(\xi_{\rm
F})).$ As a consequence, the Dirac potential reads
\bb
\label{blwformula}
V_{\rm D}(D) = {\rm tr}\left(\gamma({\cal F}_{\!D}) + {\rm ev}_{g_{\rm M}}
\left(\partial_{\!D}^{\mbox{\tiny${T^*\!\mm}\ot{\rm End}(\ep)$}}\Xi
+ \Xi^2\right)\right).
\ee
Here, respectively,
${\cal F}_{\!D}\in\Omega^2(\mm,{\rm End}(\ep))$ is the total curvature
with respect to the Dirac connection $A_{\rm D}\in{\cal A}(\xi_{\rm
F})$ and the one-form $\Xi\in\Omega^1(\mm,{\rm End}(\ep))$
measures the deviation of $A_{\rm D}$ from being a Clifford
connection. With respect to a local co-frame $(X^1,\ldots,X^n)$ on
$\mm$ this one-form reads
\bb
\Xi\,\stackrel{loc.}{=}\,
-\mbox{\small$\frac{1}{2}$}\, g_{li}\,X^l\ot\gamma(X^j)
\left([\partial_{\!D,X_j}\,,\gamma(X^i)]+
\omega^i_{\;jk}\gamma(X^k)\right),
\ee
where $(X_1,\ldots,X_n)$ is the dual frame of $(X^1,\ldots,X^n)$ and
$\omega^i_{\;jk}:=X^i(\nabla^{TM}_{\!\!X_j}X_k)$ are the corresponding
Levi-Civita connection coefficients with respect to $g_{\rm M}$ and
the chosen frame. Again, $g_{ij}\in{\cal C}^\infty(U_\alpha)$ is the
matrix element of $(g_{\rm M}(X^i,X^j))^{-1}$. Also,
``${\rm ev}_{g_{\rm M}}$'' denotes the evaluation map (contraction) with
respect to the isomorphism $\tau^*_{\rm M}\simeq \tau_{\rm M}$ of the
tangent and the cotangent bundle of $\mm$ that is provided by
$g_{\rm M}$ (c.f. \cite{AT:96} and \cite{Tol:98}).

\vspace{0.5cm}

\noindent
{\bf Remark:}\\
{\small\sl Let again, $\{(U_\alpha,\chi_\alpha)\,|\,\alpha\in\Lambda\}$ be a
family of local trivializations of a given fermion bundle
$\xi_{\rm F}.$ According to (\ref{localdops}),
$D_\alpha:=\chi_\alpha\circ D\circ\chi_\alpha^{-1}$ is fully
determined by $(\gamma_\alpha,A_\alpha,\theta_\alpha).$ Hence,
${\cal A}_{\rm D}(\xi_{\rm F})\subset{\cal A}(\xi_{\rm F})$ may
locally be identified with the set of differential forms
$\omega_\alpha\in\Omega^*(U_\alpha,{\rm End}(\cc^{\rm N}))$
which, together with $g^{ik}\in{\cal C}^\infty(U_\alpha),$ determine
$D_\alpha.$ Accordingly, the Euler-Lagrange equations}
\bb
{\cal EL}_{\rm D}(D) = 0
\ee
{\small\sl are obtained by the first variation of the
(locally defined) functional ($\Omega\subset U_\alpha,$ compact)}
\bb
{\cal S}[\gamma_\alpha,A_\alpha,\theta_\alpha]:=
\int_\Omega\!{\cal L}(\gamma_\alpha,A_\alpha,\theta_\alpha),
\ee
{\small\sl with ${\cal L}(\gamma_\alpha,A_\alpha,\theta_\alpha)
\equiv({\chi_\alpha^{-1}}^*{\cal L}_{\rm D})(D)\in\Omega^n(U_\alpha)$.
Notice, however, it can easily be
inferred from the local version of the Dirac potential
(\ref{diracpot}) that ${\cal S}={\cal S}[\gamma_\alpha,\theta_\alpha].$
Indeed, the local version of (\ref{blwformula}) reads}
\bb
\label{diracpotlocal}
V(\gamma_\alpha,A_\alpha,\theta_\alpha)&\equiv&
({\chi_\alpha^{-1}}^*V_{\rm D})(D)\\[0,2cm]\nonumber
&=& \mbox{\small$\frac{N}{2}$}\,r_{\rm M} +
\mbox{\small$\frac{1}{2}$}\,{\rm
tr}\left([\gamma_\alpha^i,\gamma_\alpha^j]
[\theta_{\alpha,i},\theta_{\alpha,j}]\right)\\[0,2cm]\nonumber
&+& \mbox{\small$\frac{1}{8}$}\,g_{ij}\,{\rm tr}\left(
\gamma_\alpha^k\,[\theta_{\alpha,k},\gamma_\alpha^i]
\gamma_\alpha^l\,[\theta_{\alpha,l},\gamma_\alpha^j]\right).
\ee
{\small\sl The notation used is as follows:\\
$\theta_\alpha\equiv X^i\ot\theta_{\alpha,i}:=
\pm{\mbox{\small$\frac{1}{n}$}}\,g_{\mu\nu}\sum_{0\leq k\leq n}\;
\sum_{1\leq i_1<i_2\cdots<i_k\leq n}X^\mu\ot
\gamma_\alpha^\nu\gamma_\alpha^{i_1}\gamma_\alpha^{i_2}\cdots
\gamma_\alpha^{i_k}\ot\theta_{i_1 i_2\cdots i_k},$ with the
abbreviation $\gamma_\alpha^i:=\gamma_\alpha(X^i)\equiv
\chi_\alpha\circ\gamma(X^i)\circ\chi_\alpha^{-1}.$ Moreover,
$r_{\rm M}\in{\cal C}^\infty(\mm)$ denotes the scalar curvature of
$\mm$ with regard to $g_{\rm M}.$}\\

\vspace{0.5cm}

It follows that Einstein's field equation of gravity is an
integral part of the Euler-Lagrange equations of Dirac type gauge
theories. In particular, the ``energy-momentum tensor'' is
specified by the Dirac type operator in question (i.e. locally
fixed by the
one-form $\theta\in\Omega^1(U,{\rm End}(\ep))\,$).\\

In the next section we discuss a specific class of Dirac type
operators which is distinguished by its Lichnerowicz
decomposition (c.f. \cite{Lich:63}). Moreover, it is shown that,
as a solution of the Euler-Lagrange equations, these Dirac type
operators spontaneously break the gauge symmetry.\\

\section[Dirac Operators of Simple Type and Symmetry Breaking]{Simple
Type Dirac Operators and Spontaneous Symmetry Breaking} In what
follows we discuss a specific class of Dirac type gauge theories.
The main feature of this class consists of permitting us to naturally
include the notion of ``spontaneous symmetry breaking'' in the
realm of Dirac type gauge theories. Eventually, we will show that the
``Yukawa coupling'' of the fermions, together with gravity, induces
spontaneous symmetry breaking without use of a
``Higgs potential''. The inner geometry of $\mm$ (i.e. of space-time
in the case of $(n,s)=(4,\mp 2)$) in the ``ground state'' of the gauge
theory is fully determined (up to boundary conditions) by the
``fermionic masses''. Here, the latter are shown to correspond to
the spectrum of a certain Hermitian section of the bundle ${\rm
End}(\xi_{\rm F}).$ Because this spectrum turns out to be constant
over $\mm$ one may thus
decompose the fermion bundle $\xi_{\rm F}$ into the Whitney sum of
the appropriate eigenbundles of the ``fermionic mass operator''
that is induced by spontaneous symmetry breaking. If the spectrum
is non-degenerated (like in the case of the Standard Model) the
eigenbundles are Hermitian line bundles which one may consider to
geometrically model
``asymptotically free fermions''.\\

Let $\xi_{\rm F}$ be a chiral fermion bundle with respect
to some Dirac triple $({\rm G},\rho_{\rm F},D).$

\begin{definition}
A Dirac type operator $D'\in{\cal D}(\xi_{\rm F})$ is called of
``simple type'' if the Bochner-Laplacian of $D'$ is defined by a
Clifford connection, i.e. ${\hat A}'_{\rm D}\in {\cal A}_{\rm
Cl}(\xi_{\rm F})\subset{\cal A}(\xi_{\rm F}).$
\end{definition}

We denote the corresponding covariant derivative again by $\dda.$
Then, the covariant derivative of the Dirac connection $A'_{\rm
D}\in{\cal A}(\xi_{\rm F})$ reads
\bb
\partial_{\!D'} = \dda + \varpi'_{\rm D}
\ee
with a unique one-form $\varpi'_{\rm D}\in\Omega^1(\mm,{\rm End}(\ep)).$
The next Proposition permits us to characterize the Dirac forms of simple
type Dirac operators of arbitrary signature.

\begin{proposition}
\label{propsimpletypedop} A Dirac type operator $D'\in{\cal
D}(\xi_{\rm F})$ is of simple type if and only if it reads \bb D'
= \ddda + \gamma_{\rm M}\ot\phi, \ee with $\phi\in\Gamma({\rm
End}^-_{\rm Cl}(\xi_{\rm F})).$
\end{proposition}

\noindent {\bf Proof:} The proof of the statement is lengthy and
somewhat technical though elementary. It is similar to the proof
already presented in \cite{AT:96} for the special case $s=n.$ A
detailed proof for arbitrary signature $s$ can be found in the
Appendix.\hfill$\Box$

\vspace{0,5cm}

Note that a simple type Dirac operator is fully determined by a
Clifford connection in the case where $\xi_{\rm F}$ is not chiral
and thus has a vanishing Dirac form. In general, however, the
Dirac connection of a simple type Dirac operator is given by a
unique Clifford connection $A\in{\cal A}_{\rm Cl}(\xi_{\rm F})$
together with the specific Dirac form
\bb
\varpi'_{\!\mbox{\tiny D}} = \Theta\wedge(\gamma_{\rm M}\ot\phi).
\ee

With respect to a local trivialization $(U_\alpha,\chi_\alpha)$ of
$\xi_{\rm F}$ the Dirac form is determined by \bb
\theta_\alpha=\pm{\mbox{\small$\frac{1}{n}$}}\,g_{ij}\,
X^i\ot\gamma_\alpha^j\gamma_{\rm M}\ot\phi_\alpha, \ee with
$1\ot\phi_\alpha:=\chi_\alpha\circ(1\ot\phi)\circ\chi_\alpha^{-1}\in
{\cal C}^\infty(U_\alpha,{\rm End}^-(\cc^{{\rm N}_{\rm F}}))$ and
$\theta_\alpha\equiv{\chi^{-1}_\alpha}^*\!\varpi'_{\!\mbox{\tiny D}}.$\\

Dirac operators of simple type define the largest class of Dirac
type operators with the corresponding Bochner-Laplace operators
defined by Clifford connections. Of course, the most important
sub-class of Dirac type operators is given by $D'=\ddda.$
They correspond to ``twisted Spin-Dirac operators'' in the case
where $\mm$ denotes a spin manifold. Notice that in the elliptic
case, Dirac operators of simple type turn out to be of importance
in the discussion of the family index theorem (c.f. \cite{Bis:86},
\cite{Qui:85}). They are also known to play a fundamental role in
the description of the minimal Standard Model within
the realm of non-commutative geometry (please see, for example, the
corresponding references already cited in the introduction). This kind
of first order differential operator is thus well-known in
physics (please, see below), as well as in mathematics. However,
in this paper we discuss them from a purely geometrical perspective
of gauge theories.\\

We turn now to the discussion of spontaneous symmetry breaking
within the realm of the presented geometrical frame. For this let
again $\xi_{\rm F}$ be a chiral fermion bundle with respect to
$({\rm G},\rho_{\rm F},D)$ where $D$ is of simple type.

\begin{proposition}
Let $D$ be a global solution of the Euler-Lagrange equation
\bb
\label{simpletypeeulergl}
{\cal EL}_{\rm D}(D)=0
\ee
such that ${\cal G}_{\rm YM}$ acts transitively on the image
of $D-\ddd_{\!\!\!A}.$ Then there exists a constant (skew-Hermitian)
section ${\cal D}\in \Gamma({\rm End}^-_{\rm Cl}(\xi_{\rm F}))$
such that $(\mm,g_{\rm M})$ is an Einstein manifold with the scalar
curvature given by
\bb
r_{\rm M}
= \lambda\,\|{\rm M}_{\rm F}\|^2.
\ee
Here, $\|{\rm M}_{\rm F}\|^2 \equiv
{\rm tr}({\rm M}_{\rm F}^\dagger{\rm M}_{\rm F})$ with
$i{\rm M}_{\rm F}:=\gamma_{\rm M}\ot{\cal D}$ representing the
``total fermionic mass operator''; $\lambda\in\rr$ is an
appropriate non-zero constant which may also depend on a suitable
normalization of ${\cal L}_{\rm D}(D).$
\end{proposition}

\noindent {\bf Proof:} The Dirac-Lagrangian of a simple type Dirac
operator reads
\bb
{\cal L}_{\rm D}(D) =
2^k({\rm N}_{\rm F}\,r_{\rm M} + {\rm tr}\phi^2)\mu_{\rm M}.
\ee

We remark that this Lagrangian depends on the connection
that is defined only with respect to $(g_{\rm M},\phi).$ Moreover,
the Euler-Lagrange equation concerning
$\phi\in\Gamma({\rm End}^-_{\rm Cl}(\xi_{\rm F}))$ is trivial.
Whence, one may conclude that a global solution of (\ref{simpletypeeulergl})
yields: $D = \ddda$ with $A\in{\cal A}_{\rm Cl}(\xi_{\rm F})$ arbitrary
and $(\mm,g_{\rm M})$ Ricci flat. However, there is actually a
bigger class of solutions of (\ref{simpletypeeulergl}). Since the
latter does not provide any dynamical condition on the sections
$\phi$ one may treat the latter as ``background fields'', similar
to the metric in the case of pure Yang-Mills gauge theory. The Euler-Lagrange
equations with respect to the corresponding Dirac-Lagrangian then
reduces to the Einstein equation
\bb
\label{simpletypeeinsteineq}
{\rm Ric}(g_{\rm M}) =
\lambda_{\rm gr}\,{\rm tr}\phi^2\,{\rm id}_{\mbox{\tiny TM}}
\ee
with $\lambda_{\rm gr}\in\rr$ being some non-zero constant which
also depends on the chosen normalization of ${\cal L}_{\rm D}(D).$
It also takes into account the appropriate physical (length)
dimension, where $\phi$ is accordingly re-scaled. The section
${\rm Ric}\in\Gamma({\rm End}(\tau_{\rm M}))$ denotes the Ricci
tensor with respect to $g_{\rm M}.$ From the Einstein
equation it follows that ${\rm d}({\rm tr}\phi^2)=0.$ Whence,
the Dirac-Lagrangian (\ref{simpletypeeulergl}) reduces to the
Einstein-Hilbert Lagrangian with ``cosmological constant'' included.
However, this constant is generated by a section
$\phi\in\Gamma({\rm End}_{\rm Cl}^-(\xi_{\rm F}))$ subject to the
condition that
$\|\phi\|^2:=<\!\!\phi,\phi\!\!>\equiv{\rm tr}(\phi^\dagger\phi)$
must be constant. Note that $\phi^\dagger = \pm\phi,$ depending on
whether $D$ is supposed to be Hermitian or skew-Hermitian.
The basic idea then is to make a polar
decomposition $\phi\stackrel{loc.}{=}\rho_{\rm F}(g)\circ{\cal D}
\circ \rho_{\rm F}(g)^{-1}$ with ${\cal D}$ being a fixed vector
of the same length as $\phi.$ To make this more precise let
${\cal W}:={\rm End}_{\rm Cl}^-(\xi_{\rm F})$ be the Hermitian
vector bundle of (complex) rank ${\rm N}_{\rm F}^2$ with total
space $W:={\rm End}_{\rm Cl}^-(\ep).$ Accordingly, let ${\goth
P}:=(P,\mm,\pi_{\rm P})$ be the frame bundle associated with
${\cal W}.$ Also, let ${\goth{E}}:=(E,\mm,\pi_{\rm E})$ be the
associated Hermitian vector bundle with total space defined by
$E:=P\times_{\rm G}{\rm End}(\cc^{{\rm N}_{\rm F}}).$ Then, by
construction $\goth{E}\simeq{\cal W}$ and we do not distinguish
between these two vector bundles. In particular, we may write
$W\ni{\goth{Z}}=[(p,\goth{z})].$ Equivalently, if $\phi\not=0$
we may consider the normalized section
$\varphi:=\phi/\|\phi\|\in\Gamma({\cal S})$ with ${\cal
S}\subset{\cal W}$ being the sphere sub-bundle. According to the
identification $\goth{E}\simeq{\cal W}$ any section $\varphi$
corresponds to a G-equivariant mapping
${\tilde\varphi}:\,P\rightarrow{\rm S}^{{\rm N'}-1}$ (${\rm
N'}=2{\rm N}_{\rm F}^2$), such that $\varphi(x) =
[(p,{\tilde\varphi}(p)]|_{p\in\pi_{\rm P}^{-1}(x)}.$ By assumption,
G acts transitively on ${\rm im}({\tilde\varphi})\subset{\rm S}^{{\rm N'}-1}.$
Hence, for arbitrarily chosen ${\goth{z}_0}\in{\rm im}({\tilde\varphi})$
we may identify the orbit of ${\goth{z}_0},$
${\rm orbit}({\goth{z}_0}),$ with ${\rm im}({\tilde\varphi}).$ Let
${\rm I}({\goth{z}_0})\subset\rho_{\rm F}({\rm G})$ be the
isotropy group of ${\goth{z}_0}.$ The mapping
\bb
\nu_\phi:\,P&\longrightarrow&{\rm orbit}(\goth{z}_0)\cr
p &\mapsto& \rho_{\rm F}(g){\goth{z}_0}\rho_{\rm F}(g^{-1}),
\ee
defines an ``H-reduction'' $({\cal Q}_\phi,\iota_\phi)$ of ${\goth P}$
with $g\in{\rm G}$ being determined (modulo ${\rm I}({\goth{z}_0})$)
by the relation ${\tilde\varphi}(pg)={\goth{z}_0}.$ Indeed, the
corresponding section
\bb
{\cal V}_\phi:\,\mm &\longrightarrow& P\times_{\rm G}{\rm G}/{\rm H}\cr
x &\mapsto& [(p,\nu_\phi(p)]|_{p\in\pi_{\rm P}^{-1}(x)}
\ee
is known to be equivalent to a specific principal
H-bundle ${\cal Q}_\phi\equiv(Q,\mm,\pi_{\rm Q},{\rm H})$ together
with an equivariant embedding $\iota_\phi:\,
{\cal Q}_\phi\hookrightarrow{\cal P}$ of principal bundles (c.f.,
for example, in \cite{KN:96}). For ``bundle reduction'' in the context
of Yang-Mills-Higgs gauge theories see also, for example,
in \cite{CW:89}, \cite{Ster:95} and \cite{Trau:80}. Here,
${\rm H}\subset{\rm G}$ is the unique sub-group equivalent to
${\rm I}({\goth{z}_0}),$ and thus
${\rm orbit}(\goth{z}_0)\simeq {\rm G/H}.$ Finally, we may define
${\cal D}\in\Gamma({\rm End}_{\rm Cl}^-(\xi_{\rm F}))$ by the section
\bb
{\cal D}:\,\mm &\longrightarrow& E\cr x &\mapsto&
[(\iota_\phi(q),{\goth{z}_0})]|_{q\in\pi_{\rm Q}^{-1}(x)}.
\ee

Of course, the section ${\cal D}$ also gives rise to an
(equivalent) H-reduction $({\cal Q},\iota)$ of ${\goth P}$ which
may be identified with $({\cal Q}_\phi,\iota_\phi)$ by ${\rm
H}\simeq{\rm I}({\tilde g}{\goth{z}_0}{\tilde g}^{-1}).$ Here,
${\tilde g}\in{\rm G}$ is determined (up to ${\rm
I}({\goth{z}_0})$) by a choice of $q_0\in Q_\phi$ and the
corresponding relation
$\varphi(\iota_\phi(q_0))\equiv{\tilde{\goth{z}}}_0=: {\tilde
g}{\goth{z}_0}{\tilde g}^{-1}.$ The rest of the statement is a
direct consequence of the Einstein equation.\hfill$\Box$

\vspace{0,5cm}

A simple type Dirac operator $D$ is said to be in the ``unitary
gauge'' provided it reads
\bb
\label{unitarygaugecondition}
D = \ddda + \gamma_{\rm M}\ot{\cal D}.
\ee

A necessary condition for the existence of the unitary gauge is that
$D-\ddda\not=0.$ If ${\cal G}_{\rm YM}$ acts transitively on the image
of the latter operator, this condition is also sufficient. A simple
type Dirac operator in the unitary gauge spontaneously breaks the Yang-Mills
gauge symmetry since in general
\bb
{\cal H}_{\rm YM}:=\{g\in{\cal G}_{\rm YM}\,|[{\cal D},g]=0\}
\ee
is a proper sub-group of the Yang-Mills gauge group
${\cal G}_{\rm YM}\subset{\cal G}_{\rm F}.$ In this case,
the Lagrangian ${\cal L}_{\rm D}(D)$ is said to
define a ``spontaneously broken fermionic gauge theory''. Note
that in the case where ${\cal G}_{\rm YM}$ acts transitively on the
sphere sub-bundle ${\cal S}\subset{\rm End}_{\rm Cl}(\xi_{\rm F})$ any
global solution of (\ref{simpletypeeulergl}) satisfying $D - \ddda \not=0$
defines a spontaneously broken fermionic gauge theory.

\vspace{0.5cm}

\noindent
{\bf Remark:}\\
{\small\sl The notion of unitary gauge and its existence is similar to that
presented in \cite{Tol:03a} (Prop. 3.2) in the case of
rotationally symmetric Higgs potentials. However, the ``mass
term'' $\|\phi\|^2$ in the Lagrangian of a
simple type Dirac operator itself does not break the symmetry, of
course. The symmetry breaking is caused by assuming that the fermionic
mass generates a non-trivial geometry. Indeed, the geometry is fully
determined by the spectrum of the (square of the) fermionic mass operator
${\rm M}^2_{\rm F}\in\Gamma({\rm End}(\xi_{\rm F}))$. Also, since the spectrum
${\rm spec}({\rm M}_{\rm F}^2)$ is constant throughout $\mm,$ one
may decompose the fermion bundle into the Whitney sum of the
corresponding eigenbundles of ${\rm M}_{\rm F}^2,$ i.e.}
\bb
\label{asymptoticfreefermions}
\xi_{\rm F} &=&
\bigoplus_{\mbox{\tiny${\rm m}^2\in{\rm spec}({\rm M}_{\rm F}^2)$}}
\xi_{{\rm F},{\rm m}^2}\nonumber\\[0,2cm]
&=&
{\rm ker}({\rm M}_{\rm F}^2) \op \left[
\bigoplus_{\mbox{\tiny${\rm m}^2\in
{\rm spec}({\rm M}_{\rm F}^2)\backslash\{0\}$}} \xi_{{\rm F},{\rm m}^2}\right].
\ee

{\small\sl The total curvature on $\xi_{\rm F}$ with respect to a simple type
Dirac operator satisfying (\ref{simpletypeeulergl}) is given by}
\bb
\label{gravindcurvdecomp}
{\cal F}_{\!D} = {/\!\!\!\!R}+F_{\!A}+{\rm M}_{\rm F}^2\,\Theta\wedge\Theta
-\dda^{\!\!\mbox{\tiny${\rm End}(\ep)$}}{\rm M}_{\rm F}\wedge\Theta.
\ee
{\small\sl Here, respectively,
${/\!\!\!\!R}\in\Omega^2(\mm,{\rm End}(\ep))$ is the lifted
(semi-)Riemannian curvature with respect to $g_{\rm M},$ and
$F_{\!A}\equiv{\cal F}_{\!\!\mbox{\tiny$\ddda$}}-{/\!\!\!\!R}
\in\Omega^2(\mm,{\rm End}(\ep))$ is the ``twisting curvature'' with respect to
the Clifford connection $A\in{\cal A}_{\rm Cl}(\xi_{\rm F})$ that
is determined by $D.$ In contrast to ${/\!\!\!\!R},$ which is
determined by the spectrum of ${\rm M}_{\rm F}^2,$ the twisting
curvature $F_{\!A}$ is completely arbitrary. For this reason it
is natural to assume that $A$ is purely topological, i.e. flat. In
this case, the curvature of $\xi_{\rm F}$ is fully determined by
the spectrum of the fermionic mass operator. As a consequence, for
$n=4$ the chiral fermion bundle must indeed be real. If in addition
$\mm$ is a spin-manifold, then
$\xi_{{\rm F},{\rm m}^2}\simeq\tau_{\rm spin}\ot_\mm
\zeta_{{\rm F},{\rm m}^2},$ where the latter is an Hermitian line
bundles if and only if ${\rm spec}({\rm M}_{\rm F}^2)\backslash\{0\}$ is
non-degenerated. Consequently, when restricted to the residual
group H, the fermionic representation $\rho_{\rm F}$ decomposes
into the sum of the trivial representation and irreducible
U(1)-representations\footnote{To date, electromagnetism is the only
Abelian gauge theory that is physically well-established.
Moreover, as a matter of fact massless but electrically
charged particles are unknown in nature.}. The latter are
either trivial, and hence $\xi_{{\rm F},{\rm m}^2}$ corresponds
to electrically uncharged but massive fermion or, for non-trivial
representations, $\xi_{{\rm F},{\rm m}^2}$ corresponds to a
massive electrically charged particles. Apparently, together
with spin, the assumption that the Clifford connection $A$ is
flat imposes crucial restrictions on the fermion bundle. In fact,
in this case (up to algebraic torsion)
$\xi_{{\rm F}}\simeq\bigoplus_{k=1}^{{\rm N}_{\rm F}}\tau_{\rm spin}.$
Note that, if $n=4$ and ${\rm spec}({\rm M}_{\rm F}^2)$ is
non-degenerate, the existence of a flat Clifford connection on
$\xi_{{\rm F}}$ (again, up to torsion) becomes equivalent to
the reality of the latter.}

\vspace{0.5cm}

\begin{definition}
A fermion bundle $\xi_{\rm F}$ is said to be in the ``unitary
gauge'' provided it is defined with respect to a Dirac triple
$({\rm G},\rho_{\rm F},D)$ such that $D$ is in the unitary gauge.
More generally, a fermion bundle is called ``massive'' if it is
gauge equivalent to a fermion bundle in the unitary gauge. The
corresponding element of ${\cal G}_{\rm YM}\subset{\cal G}_{\rm F}$
is referred to as a ``unitary gauge transformation''.
\end{definition}

On a massive fermion bundle there exists a distinguished class
of connections.

\begin{definition}
\label{reducibleconnection} A connection $A\in{\cal A}(\xi_{\rm
F})$ on a massive fermion bundle $\xi_{\rm F}$ is called
compatible with $D$ provided the corresponding covariant
derivative $\nabla^\ep$ commutes with the appropriate total
fermionic mass operator. That is,
\bb
\label{reducibleconnectionrel}
\nabla^{{\rm End}(\ep)}_{\!\!X}{\rm M}_{\rm F} = 0
\ee
for all smooth tangent vector fields $X\in\Gamma(\tau_{\rm M}).$
\end{definition}

This definition expresses the H-reducibility of a connection on
$\xi_{\rm F}$ in terms of Dirac type operators which spontaneously
break the gauge symmetry. The Definition
(\ref{reducibleconnection}) is in fact analogous to the Definition
2.1 in \cite{Tol:03a} for a spontaneously broken Yang-Mills-Higgs
gauge theory. Note that (\ref{reducibleconnectionrel}) is
equivalent to the condition
\bb
D'\circ {\rm M}_{\rm F} = -{\rm M}_{\rm F}\circ D',
\ee
with $D'\in{\cal D}(\xi_{\rm F})$ being identified with
$\gamma\circ\nabla^\ep.$ In particular,
one may assume that the Clifford connection which defines the
Bochner-Laplacian of $D=\ddda + i{\rm M}_{\rm F}$ is compatible
with the latter. This holds true if and only if
\bb
\label{dalambertop}
D^2 = \ddda^2 - {\rm M}_{\rm F}^2.
\ee

Hence, the Clifford connection of the Bochner-Laplacian
$\triangle_{\mbox{\tiny D}}$ is compatible with spontaneous
symmetry breaking if and only if ``the square of the sum equals
the sum of the squares''. We note that, from a geometrical point
of view, it is the condition
$\nabla^{{\rm End}(\ep)}{\rm M}_{\rm F} \not= 0$ that yields
``massive vector bosons'' (please see below). In other words,
the existence of a non-trivial ``Yang-Mills mass operator'' can be
expressed by the violation of the compatibility condition
(\ref{dalambertop}).\\

\begin{definition}
We call a simple type Dirac operator $D$ to define a
``(semi-classical) fermionic vacuum'' if $D$ is gauge equivalent
to $\ddda + i{\rm M}_{\rm F}$ where the corresponding Clifford
connection $A\in{\cal A}(\xi_{\rm F})$ is purely topological. In
this case, $D$ in the unitary gauge is denoted by
\bb
\label{fermionicvacuum}
 \ddd_{\!{\cal D}}\equiv\ddd + i{\rm M}_{\rm F}.
 \ee
\end{definition}

Clearly, when restricted to the appropriate eigenbundles this
operator corresponds to Dirac's well-known first order
differential operator $i\ddd - {\rm m}$ and thus provides us with
the appropriate physical interpretation of
${\rm spec}({\rm M}^2_{\rm F})$ (and hence also with ${\cal D}$).
For example, in the case of
$(n,s)=(4,\mp2)$ there is always a local frame such that the total
symbol $\sigma(i\ddd_{\!{\cal D}})$ coincides with the principal
symbol of (\ref{fermionicvacuum}). Every time-like $\xi\in
T^*\mm\subset{\rm End}(\ep)$ and eigenvector ${\goth z}\in\ep$ of
${\rm M}_{\rm F}^2 $ (with eigenvalue ${\rm m}^2$) yields
$\sigma(i\ddd)(\xi){\goth z}\stackrel{loc}{=}
\gamma(\xi){\goth z}=\pm{\rm m}{\goth z}.$ Hence, one obtains the
usual relation between momentum and mass:
$g_{\rm M}(\xi,\xi)=\pm{\rm m}^2$ of a point-like particle.\\

From a geometrical point of view a ``fermionic vacuum'' may be
regarded as a fermion bundle $\xi_{\rm F,red}:=
(\ep_{\rm red},\mm,\pi_{\ep,{\rm red}})$ with respect to the Dirac
triple $({\rm H},\rho_{\rm F,red}, \ddd_{\!{\cal D}}).$ Here,
respectively, $\ep_{\rm red}:=
Q\times_{\rm H}{\cc^{2^{\rm k}}\!\ot\cc^{{\rm N}_{\rm F}}}$ and
$\rho_{\rm F,red}:=\rho_{\rm F}|_{\rm H}.$ Notice that
$\xi_{\rm F}\simeq\xi_{\rm F,red}$ via the bundle mapping
$[(q,z)]\mapsto[(\iota(q),z)].$ Accordingly, we shall not
distinguish between these two bundles and proceed to say that a fermion
bundle $\xi_{\rm F}$ can be generated from a fermionic vacuum if
it is determined by a Dirac triple of the form
$({\rm H},\rho_{\rm F,red}, \ddd_{\!{\cal D}}).$ In other words,
$\xi_{\rm F}$ is
generated from a fermionic vacuum provided the corresponding frame
bundle ${\cal P}$ can be considered as a prolongation of the frame
bundle ${\cal Q}$ that corresponds to some fermion bundle
$\xi_{\rm F,red}.$ Finally, the Dirac potential of a fermionic
vacuum has the particular simple form
\bb
V_{\rm D}(\ddd_{\!{\cal
D}}) &=& \frac{\lambda}{2}\,{\mbox{\small$<\!{\rm M}^2_{\rm F}\!>$}},
\ee
where ${\mbox{\small$<\!{\rm M}^2_{\rm F}\!>$}}:=
{\mbox{\small$\frac{1}{{\rm N}_{\rm F}}$}}
\sum_{a=1}^{\mbox{\tiny${\rm N}_{\rm F}$}}{\rm m}_a^2$ and
$\lambda\in\rr$ is a suitable non-zero constant.\\

The idea of a fermionic vacuum is mainly motivated by a
geometrical description of perturbation theory used in quantum
field theory. As already mentioned above the fermion bundle
$\xi_{\rm F}$ is considered as a ``perturbation'' of a fermionic
vacuum $\xi_{\rm F,red}.$ Such a perturbation cannot
change the topology of $\xi_{\rm F}$ but its geometry.
The notion of a fermionic vacuum itself
puts severe topological restrictions on a fermion
bundle\footnote{One might speculate that
``quantum fluctuations'' will lead to a
change of the topology of the fermionic vacuum for it basically
adds ``quantum corrections'' to the fermionic mass spectrum.}.
Before we explain this in more detail, however, we shall discuss in the
next section a more specific class of simple type Dirac operators
which takes into account that, within the Standard Model of Particle
Physics, the Higgs boson is described by a sub-representation of
$\rho_{\rm F}$ instead of the fundamental representation.
Moreover, we shall discuss the need of ``fermionic doubling'' and
the fermionic Lagrangian within the presented setup.

\section[Dirac-Yukawa Operators and Fermionic Lagrangian]{Dirac-Yukawa Type
Operators and the Fermionic Lagrangian}

In the last section we discussed a distinguished class of Dirac
type operators on a fermion bundle. Their basic feature is to give
rise to a reduction of the underlying gauge symmetry. Moreover,
these Dirac type operators also determine a distinguished class of
connections on the fermion bundle. In the next two sections we
specialize the presented frame in order to geometrically describe
the action of the Standard Model of Particle Physics in terms of a
specific Dirac-Lagrangian. For this, we first discuss a certain
``refinement'' of simple type Dirac operators which will then be
called ``Dirac-Yukawa operators''. In what follows, we also
discuss an important consequence of the occurrence of the grading
involution $\gamma_{\rm M}$ in the definition of simple type Dirac
operators. This turns out to parallel the occurrence of this grading
involution in A. Connes' non-commutative geometry (c.f., for example,
in \cite{Con:94}, \cite{GIS:97}, \cite{GV:93}, \cite{KS:96},
\cite{LMMS:96}, \cite{LMMS:97} and \cite{SZ:95}).

\subsection{Yukawa Bundles and Dirac Operators of Yukawa Type}
To start with, let again $\xi_{\rm F}$ be a chiral fermion bundle
with respect to $({\rm G},\rho_{\rm F},D),$ where $D$ is of simple
type. Also let $\xi_{\rm H}\subset\xi_{\rm F}$ be a sub-vector
bundle of rank ${\rm N}_{\rm H}<{\rm N}_{\rm F}$ on which
$\tau_{\rm Cl}$ acts trivially. We denote its dual by $\xi_{\rm
H}^*.$ The structure group of $\xi_{\rm H}$ is a specific
sub-group of $\rho_{\rm F}({\rm G}).$ It will be denoted by
$\rho_{\rm H}({\rm G}).$ The gauge group of $\xi_{\rm H}$ is
accordingly denoted by ${\cal G}_{\rm H}\subset{\cal G}_{\rm YM}
\subset{\rm Aut}(\xi_{\rm
 H})\subset{\rm Aut}(\xi_{\rm F})$
(the bundle automorphisms of $\xi_{\rm H}$ over the identity on $\mm.$)

\begin{definition}
\label{yukawamapping} Let $E_{\rm H}\subset\ep$ be the total space
of $\xi_{\rm H},$ and let $\pi_{\rm H}$ be the appropriate projection
mapping onto the base manifold $\mm.$ Also, let again $W:={\rm
End}^-_{\rm Cl}(\ep).$ We call the sub-vector bundle $\xi_{\rm
Y}\subset\xi_{H}^*\ot_\mm\xi_{\rm W}$ the ``Yukawa bundle'' (with
respect to the above data) if its structure group acts as
follows: For each $h\in{\rm Aut}(E_{\rm H})$ there is a unique
$g\in{\rm Aut}^+_{\rm Cl}(\ep)$ such that ${\cal
Y}({h^{-1}\goth{z}})={\rm Ad}_{g^{-1}}({\cal Y}({\goth{z}}))$ for
all ${\goth{z}}\in E_{\rm H}$ and ${\cal Y}\in E_{\rm H}^*\ot W.$
In this case we call $\xi_{\rm H}$ the ``Higgs bundle'' (again,
with respect to the above given data). A section ${\cal
Y}\in\Gamma(\xi_{\rm Y})$ of the Yukawa bundle is called a
``Yukawa mapping'' provided that it fulfills the following conditions:
Considered as a bundle mapping the Yukawa mapping ${\cal Y}$ is
injective and anti-Hermitian, i.e. ${\cal Y}({\goth{z}})^\dagger =
-{\cal Y}({\goth{z}})$ for all ${\goth{z}}\in E_{\rm H}.$ Moreover,
we assume that it satisfies the requirement
${\cal Y}(\partial_{\!A,X}\varphi)=[\partial_{\!A,X},{\cal Y}(\varphi)]$
for all Clifford connections on $\xi_{\rm F}$ (and thus for all
induced connections on $\xi_{\rm H}$), sections
$\varphi\in\Gamma(\xi_{\rm H})$ and tangent vector fields
$X\in\Gamma(\tau_{\rm M}).$
\end{definition}

Note that for each connection on $\xi_{\rm F}$ with covariant
derivative $\nabla^\ep,$ the operator $[\nabla^\ep_{\!\!\!X},{\cal
Y}(\varphi)]- {\cal Y}(\nabla^{E_{\rm H}}_{\!\!\!X}\varphi)$ on
the fermion bundle $\xi_{\rm F}$ defines a connection on $\xi_{\rm
H}^*\ot\xi_{\rm W}$ with the covariant derivative $\nabla^{E^*_{\rm
H}\ot W}\equiv\nabla^{E^*_{\rm H}}\ot 1 + 1\ot\nabla^W.$ Hence, a
Yukawa mapping is assumed to be covariantly constant with respect
to any Clifford connection. By the definition of the Yukawa bundle
it then follows that a Yukawa mapping has to be a constant
section. For instance, in the case of the Standard Model the
Yukawa mapping (\ref{yukawamapping}) is parameterized by the
``Yukawa coupling constants''. The representations $\rho_{\rm H}$
and $\rho_{\rm F}$ are known to be related by the
``hyper-charges'' of the fermions and the Higgs boson.

\begin{definition}
We call a Dirac type operator $D$ on a fermion bundle $\xi_{\rm
F}$ a ``Dirac-Yukawa operator'' if there is a section of the Higgs
bundle,
 $\varphi\in\Gamma(\xi_{\rm H}),$ such that
\bb \label{diracyukawaop} D = \ddda + \gamma_{\rm M}\ot{\cal
Y}(\varphi). \ee According to its physical interpretation we call
the section ${\cal Y}(\varphi)\in\Gamma({\rm End}^-_{\rm
Cl}(\xi_{\rm F}))$ the ``Yukawa coupling term'' with respect to
$({\cal Y},\varphi)\in\Gamma(\xi_{\rm Y}\times_\mm\xi_{\rm H}).$
\end{definition}

A Yukawa mapping defines an additional data on a fermion bundle
which in some sense is not natural within the frame of Dirac type
gauge theories. For this reason we shall refer to the data $({\rm
G},\rho_{\rm F},D),$ with $D$ being a Dirac-Yukawa operator, as a
``Dirac-Yukawa model''. A necessary condition for a Dirac-Yukawa
operator to spontaneously break the underlying gauge symmetry is
that $\varphi\in\Gamma(\xi_{\rm H})$ does not vanish. Again, this
condition is also sufficient provided G acts transitively on the
image of the section ${\cal Y}(\varphi).$ Assuming this is the case
it follows from the definition of the Higgs bundle and the Yukawa
mapping that there must exist a constant section
${\cal V}\in\Gamma(\xi_{\rm H})\backslash\{{\cal O}\}$ (with ${\cal O}$
being the zero-section) such that in the unitary gauge
\bb
D = \ddda
+ \gamma_{\rm M}\ot{\cal Y}({\cal V}).
\ee

Analogous to our previous definition we consider a Dirac-Yukawa
operator to define a (semi-classical) fermionic vacuum if it is
gauge equivalent to $\ddd_{\!\cal V}\equiv\ddd + i{\rm M}_{\rm F}$
with the total fermionic mass operator $i{\rm M}_{\rm
F}:=\gamma_{\rm M}\ot{\cal Y}({\cal V}).$ Notice that the spectrum
of the total fermionic mass operator is independent of
the choice of $\goth{Z}_0\in{\rm End}(\cc^{{\rm N}_{\rm F}}).$
This reduces to $\goth{Z}_0={\rm G}_{\rm Y}({\bf z}_0)$ in the
case where the gauge symmetry is spontaneously broken by a
Dirac-Yukawa operator. Here, ${\rm G}_{\rm Y}\in
{\rm Hom}(\cc^{{\rm N}_{\rm H}},{\rm End}(\cc^{{\rm N}_{\rm F}}))$
is the matrix of the ``Yukawa coupling constants'' and
${\bf z}_0\in\cc^{{\rm N}_{\rm H}}.$ In particular, we obtain
${\rm orbit}({\goth Z}_0)={\rm G}_{\rm Y}({\rm orbit}({\bf z}_0)).$
Hence, from the properties of the Yukawa mapping it can be inferred
that the ``little group'' ${\rm H}\subset{\rm G}$ crucially depends
on $\rho_{\rm H}\subset\rho_{\rm F}.$

\subsection{The Fermionic Lagrangian}
Next, we discuss the fermionic Lagrangian within the presented frame.
By definition, the grading involution of a chiral fermion bundle
$\xi_{\rm F}=\xi_{\rm F}^+\op\xi_{\rm F}^-$ reads $\Gamma =
\gamma_{\rm M}\ot\chi.$ Consequently, the total space $\ep$ of the
fermion bundle decomposes as
\bb
\label{fermiondoubling}
\ep &=&
\ep^+\op\ep^-\cr &=& (\ep_{\rm LL}\op\ep_{\rm RR})\op (\ep_{\rm
RL}\op\ep_{\rm LR})
\ee
where, respectively,
\bb
\label{fermiondecomp}
\ep_{\rm LL} &:=&\{{\goth{z}}\in\ep\,|\,(\gamma_{\rm M}\ot
1){\goth{z}}=-{\goth{z}},\, (1\ot\chi){\goth{z}}=-{\goth{z}}\},\cr
\ep_{\rm RR} &:=&\{{\goth{z}}\in\ep\,|\,(\gamma_{\rm M}\ot
1){\goth{z}}={\goth{z}},\, (1\ot\chi){\goth{z}}={\goth{z}}\},\cr
\ep_{\rm RL} &:=&\{{\goth{z}}\in\ep\,|\,(\gamma_{\rm M}\ot
1){\goth{z}}={\goth{z}},\, (1\ot\chi){\goth{z}}=-{\goth{z}}\},\cr
\ep_{\rm LR} &:=&\{{\goth{z}}\in\ep\,|\,(\gamma_{\rm M}\ot
1){\goth{z}}=-{\goth{z}},\, (1\ot\chi){\goth{z}}={\goth{z}}\}.
\ee
Let $\pi_{\rm R/L}:=\frac{1}{2}(1 \pm (\gamma_{\rm M}\ot 1))$ and
$\rho_{\rm R/L}:=\frac{1}{2}(1 \pm (1\ot \chi)).$ The appropriate
projection mappings of the respective subspaces
(\ref{fermiondecomp}) of $\ep$ are denoted by $\pi_{\rm
LL}\equiv\pi_{\rm L}\circ\rho_{\rm L}=\rho_{\rm L}\circ\pi_{\rm
L},$ $\pi_{\rm RR}\equiv\pi_{\rm R}\circ\rho_{\rm R}=\rho_{\rm
R}\circ\pi_{\rm R},$ $\pi_{\rm RL}\equiv\pi_{\rm R}\circ\rho_{\rm
L}=\rho_{\rm L}\circ\pi_{\rm R}$ and $\pi_{\rm LR}\equiv\pi_{\rm
L}\circ\rho_{\rm R}=\rho_{\rm R}\circ\pi_{\rm L}.$ Consequently,
$\pi_+ = \pi_{\rm RR} + \pi_{\rm LL}$ and $\pi_- = \pi_{\rm RL} +
\pi_{\rm LR}.$ For $\phi\in\Gamma({\rm End}_{\rm Cl}(\xi_{\rm
F}))$ we also define $1\ot\phi_{\rm LL}:=\rho_{\rm
L}\circ(1\ot\phi)\circ\rho_{\rm L} \in\Gamma({\rm End}_{\rm
Cl}(\xi_{\rm F,LL}\op\xi_{\rm F,RL}))\simeq \Gamma({\rm End}_{\rm
Cl}(\xi_{\rm F,LL}))\op \Gamma({\rm End}_{\rm Cl}(\xi_{\rm
F,RL})),$
 $1\ot\phi_{\rm RL}:=\rho_{\rm R}\circ(1\ot\phi)\circ\rho_{\rm L}
\in\Gamma({\rm Hom}_{\rm Cl}(\xi_{\rm F,LL}\op\xi_{\rm F,RL},
\xi_{\rm F,LR}\op\xi_{\rm F,RR}))\simeq \Gamma({\rm Hom}_{\rm
Cl}(\xi_{\rm F,LL},\xi_{\rm F,LR}))
\op\Gamma({\rm Hom}_{\rm Cl}(\xi_{\rm F,RL},\xi_{\rm F,RR})),$ etc.\\

If $\mm$ denotes a spin manifold, then $\ep\simeq S\ot E_{\rm F},$
where $S$ is the total space of the spinor bundle $\tau_{\rm spin}$
(with respect to some chosen spin structure) and $E_{\rm F}$ is the
total space of some Hermitian vector bundle $\zeta_{\rm F}.$ In
this case, the fermion bundle $\xi_{\rm F}\simeq\tau_{\rm
spin}\ot\zeta_{\rm F}$ is chiral if and only if $\zeta_{\rm F}$ is
$\zz_2-$graded, i.e. $E_{\rm F} = E_{\rm F,R}\op E_{\rm F,L}.$
Here, $E_{\rm F, R/L}$ are considered as the eigenspaces of $\chi$
with respect to the eigenvalues $\pm 1.$ Then, for instance,
$\ep_{\rm LL}\simeq S_{\rm L}\ot E_{\rm F,L},$ etc. Consequently,
like in non-commutative geometry, the fermionic degrees of freedoms
are doubled in the geometrical description presented here (c.f.
again the corresponding discussion in \cite{LMMS:96},
\cite{LMMS:97}). Indeed, as far as the Standard
Model is concerned only
\bb
\label{pyhssubspace}
\ep_{\rm phy}\equiv\ep^+=(\ep_{\rm LL}\op\ep_{\rm RR})
\ee
represents the ``true'' physical degrees of freedom.\\

With this in mind the ``fermionic Lagrangian'' of $D$ may be
defined as the following specific quadratic form on
$\Gamma(\xi_{\rm F})$ (taking its value in the top forms of $\mm$):
\bb
\label{fermioniclagrangian}
{\cal L}_{\rm
F}:\, {\cal D}(\xi_{\rm F}) &\longrightarrow&
\Gamma(\xi_{\rm F}\ot_\mm\Lambda^{\!\rm n}\tau^*_{\rm M})\nonumber\\[0.2cm]
D &\mapsto& \left\{
\begin{array}{ccc}
  \Gamma(\xi_{\rm F}) & \longrightarrow & \Omega^{\rm n}(\mm) \\
  \psi & \mapsto & \;<\psi,D_+\psi>_{\ep}\mu_{\rm M}.
\end{array}\right.
\ee Here, $<\!\cdot,\cdot\!>_{\ep}$ is the Hermitian product on
$\ep$ and $D_\pm\equiv\pi_\mp\circ D\circ\pi_\pm:\,
\Gamma(\xi_{\rm F}^\pm)\rightarrow\Gamma(\xi_{\rm F}^\mp)$ such
that $D\in{\cal D}(\xi_{\rm F})$ reads \bb
D = \left(%
\begin{array}{cc}
  0 & D_- \\
  D_+ & 0 \\
\end{array}%
\right):\,
\begin{array}{c}
  \Gamma(\xi_{\rm F}^+) \\
  \op \\
  \Gamma(\xi_{\rm F}^-) \\
\end{array}
\longrightarrow
\begin{array}{c}
  \Gamma(\xi_{\rm F}^+) \\
  \op \\
  \Gamma(\xi_{\rm F}^-) \\
\end{array}.
\ee

It is common use to also refer to the operators $D_{\pm}$ themselves
as Dirac type operators although the square of these operators is
usually not defined\footnote{Equivalently, if, for instance,
the operator
$D_+:\,\Gamma(\xi_{\rm F}^+)\rightarrow\Gamma(\xi_{\rm F}^-)$ is
identified by the operator
$\left(%
\begin{array}{cc}
  0 & 0 \\
  D_+ & 0 \\
\end{array}%
\right):\,\Gamma(\xi_{\rm F})\rightarrow\Gamma(\xi_{\rm F})$ it
follows that $D_+^2\equiv 0.$ Hence, it is not a Dirac type
operator in the sense presented here. However, every (anti-)
symmetric Dirac type operator $D$ is fully determined by $D_+.$}. The
Hermitian product on $\ep$ depends on the signature of $D.$ For instance,
in the respective cases of Lorentzian and Euclidean signature the
following is obtained for all $\goth{z},\goth{z}'\in\ep:$
\bb
<\!\goth{z},\goth{z}'\!>_\ep\; :=
\;\left\{%
\begin{array}{cc}
  \overline{\goth{z}_-}\;\goth{z}'_+ \;+\;
  \overline{\goth{z}_+}\;\goth{z}'_- & ({\rm Lorentzian}\;{\rm sign.}),
  \\[0.2cm]\nonumber
  \overline{\goth{z}_+}\;\goth{z}'_+ \;+\;
  \overline{\goth{z}_-}\;\goth{z}'_- & ({\rm Euclidean}\;{\rm sign.}),\\
\end{array}%
\right.
\ee
where ${\bar\goth{z}}$ means either the Dirac or
Hermitian conjugate of the ``spinor degrees'' of freedom of
$\goth{z}.$ More precisely, let $\pi:\,{\rm F}\ep\rightarrow\mm$
be the frame bundle of $\xi_{\rm F},$ such that
$\ep\ni{\goth{z}}\simeq[(p,{\bf z}=\sum_{i=1}^{2^{\rm k}}{\bf
s}_i\ot{\bf z}_i)] \in{\rm F}\ep\times_{{\rm spin(n)}
\times\rho_{\rm F}({\rm G})} \cc^{2^{\rm k}}\ot\cc^{{\rm N}_{\rm F}}.$
Then, the notation $\overline{\goth{z}_1}\,\goth{z}_2$ means:
$\overline{\goth{z}_1}\,\goth{z}_2\equiv\overline{{\bf z}_1}\,{\bf
z}_2:= \sum_{i=1}^{2^{\rm k}}(\overline{{\bf s}_{1,i}}\;{\bf s}_{2,i})
({\bf z}_{1,i}^\dagger\,{\bf z}_{2,i}).$ By the
definition of the fermion bundle, this value is clearly
independent of the choice of $p\in{\rm F}\ep$
and thus independent of the representative ${\bf z}$ of $\goth{z}.$
Hence, in the cases considered, the fermionic Lagrangian
(\ref{fermioniclagrangian}) reads
\bb
{\cal L}_{\rm F}(D)(\psi)\; :=
\;\left\{%
\begin{array}{cc}
\left(\overline{\psi_+}\;D_+\psi_+\right)\mu_{\rm M} & ({\rm
Lorentzian}\;{\rm sign.}),
\\[0.2cm]\nonumber
\left(\overline{\psi_-}\;D_+\psi_+\right)\mu_{\rm M} &
({\rm Euclidean}\; {\rm sign.}).\\
\end{array}%
\right.
\ee

The $D_+$ part of simple type Dirac operators has the form
\bb
\label{positivsimpletypedop}
D_+ &=&
\left(%
\begin{array}{cc}
  \ddda & \Phi_{\rm LR} \\
  -\Phi_{\rm RL} & \ddda \\
\end{array}%
\right)\\[0.2cm]\nonumber
&\equiv& \ddda + \gamma_{\rm M}\ot\phi_+
\ee
where, respectively,
$\Phi_{\rm LR}:=\gamma_{\rm M}\ot{\tilde\phi}_{\rm LR}\in
\Gamma({\rm Hom}(\xi_{\rm F,RR},\xi_{\rm F,RL}))$ and
$\Phi_{\rm RL}:=-\gamma_{\rm M}\ot{\tilde\phi}_{\rm RL}\in
\Gamma({\rm Hom}(\xi_{\rm F,LL},\xi_{\rm F,LR})).$ The mapping
${\tilde\phi}_{\rm LR}$ equals $\phi_{\rm LR}$ restricted to
$\Gamma({\rm Hom}_{\rm Cl}(\xi_{\rm F,LL},\xi_{\rm F,LR}))$ and
${\tilde\phi}_{\rm RL}$ equals $-\phi_{\rm RL},$ restricted to
the sub-space
$\Gamma({\rm Hom}_{\rm Cl}(\xi_{\rm F,RR},\xi_{\rm F,RL})).$
Since (\ref{positivsimpletypedop}) formally looks like a simple
type Dirac operator, we also refer to it as a Dirac operator of
simple type. For Lorentzian or Euclidean signature the corresponding
fermionic Lagrangian reads:
\bb
{\cal L}_{\rm F}(\ddda + \gamma_{\rm M}\ot\phi)(\psi)\; &=&
\left\{%
\begin{array}{cc}
  \left(\overline{\psi_+}(\ddda + \gamma_{\rm M}\ot\phi_+)\psi_+\right)\mu_{\rm M} &
  ({\rm Lorentzian}\;{\rm sign.}), \\[0.25cm]\nonumber
  \left(\overline{\psi_-}(\ddda + \gamma_{\rm M}\ot\phi_+)\psi_+\right)\mu_{\rm M} &
  ({\rm Euclidean}\;{\rm sign.}). \\
\end{array}%
\right.\\[0.35cm]\nonumber
&=&
\;\left\{%
\begin{array}{c}
\left(\overline{\psi_{\rm LL}}\;\ddda\psi_{\rm LL}\;+\;
\overline{\psi_{\rm RR}}\;\ddda\psi_{\rm RR}\right)\mu_{\rm M}\;
+\\[0.25cm]\nonumber
\left(\overline{\psi_{\rm LL}}(1\ot{\tilde\phi}_{\rm LR})\psi_{\rm
RR}\;+\; \overline{\psi_{\rm RR}}(1\ot{\tilde\phi}_{\rm
RL})\psi_{\rm LL}\right)
\mu_{\rm M},\\[0.45cm]\nonumber
\left(\overline{\psi_{\rm RL}}\;\ddda\psi_{\rm LL}\;+\;
\overline{\psi_{\rm LR}}\;\ddda\psi_{\rm RR}\right)\mu_{\rm M}\;
+\\[0.25cm]\nonumber
\left(\overline{\psi_{\rm RL}}(1\ot{\tilde\phi}_{\rm LR})\psi_{\rm
RR}\;+\; \overline{\psi_{\rm LR}}(1\ot{\tilde\phi}_{\rm
RL})\psi_{\rm LL}\right)
\mu_{\rm M}.\\
\end{array}%
\right.
\ee
\phantom{xxx}

Note that $D$ is formally self-adjoint if and only if
$D_-=D_+^\dagger.$ Also note that $\phi^\dagger=-\phi$ if and only
if $\phi_+^\dagger=-\phi_+,$ which in turn is equivalent to
$(1\ot{\tilde\phi}_{\rm RL})=(1\ot{\tilde\phi}_{\rm LR})^\dagger.$
Here, all mappings are considered to be defined on the total
space $\Gamma(\xi_{\rm F}).$ In case of $D$ being (anti-)
Hermitian we may set, respectively,
$(1\ot{\tilde\phi}):=(1\ot{\tilde\phi}_{\rm LR})$ and
${\tilde\Phi}\equiv\gamma_{\rm M}\ot{\tilde\phi}.$\\

Finally, for a Dirac-Yukawa operator one obtains
\bb
\label{diracyukawaopplus}
D_+ &=&
\left(%
\begin{array}{cc}
  \ddda & {\cal G}_{\rm Y}(\varphi) \\[0.1cm]
  -{\cal G}_{\rm Y}(\varphi)^\dagger & \ddda \\
\end{array}%
\right)\\[0.2cm]\nonumber
&\equiv& \ddda + \gamma_{\rm M}\ot{\tilde{\cal Y}}(\varphi),
\ee
with a smooth mapping
\bb
\label{yukawamappingplus}
{\cal G}_{\rm Y}:\,\Gamma(\xi_{\rm H})&\longrightarrow&
\Gamma({\rm Hom}(\xi_{\rm F,RR},\xi_{\rm F,RL}))\cr
\varphi &\mapsto&
\gamma_{\rm M}\ot{\tilde\phi}:={\cal G}_{\rm Y}(\varphi)
\ee
that is induced by an appropriate Yukawa mapping (\ref{yukawamapping})
and where $\varphi\in\Gamma(\xi_{\rm H})$ is a section of the Higgs
bundle. We may therefore formally refer to the operator
(\ref{diracyukawaopplus}) also as a Dirac-Yukawa operator.\\

As an example, we consider the fermionic Lagrangian of a
Dirac-Yukawa type operator of Lorentzian signature which
spontaneously breaks the gauge symmetry. In the case of
${\rm N}_{\rm F,L}:=2, {\rm N}_{\rm F,R}:=1$ the fermionic
Lagrangian (\ref{fermioniclagrangian}) reads
\bb
{\cal L}_{\rm F}(iD)(\psi)
\; = \;<\!\nu_{\rm L},i\ddd\nu_{\rm
L}\!>_{{\ep}_\nu}\mu_{\rm M} \;+ \;<\!e,(i\ddd -
m)e\!>_{{\ep}_e}\mu_{\rm M}, \ee with the suggestively physical
notation $\psi_{\rm LL}\equiv(\nu_{\rm L},e_{\rm L})$ and
$\psi_{\rm RR}\equiv e_{\rm R}$ for the ``state'' of the
left-handed and right-handed leptons, respectively. Here,
$\nu_{\rm L}\equiv\nu_{\rm LL}\op\nu_{\rm RL}$ and $e\equiv e_{\rm
L}\op e_{\rm R}$ are considered as eigen sections of the total
fermionic mass matrix which correspond to the eigenvalues zero
and $m\in\rr_+^{\mbox{\tiny$\times$}}.$ Physically, one may
interpret the corresponding (isomorphism class of) eigenbundles
$\xi_{\rm F}^\nu$ and $\xi_{\rm F}^e$ (with $\xi_{\rm
F}\simeq\xi_{\rm F}^\nu\op\xi_{\rm F}^e$) as
``asymptotically free particles''.

\vspace{0.5cm}

\noindent
{\bf Remark:}\\
{\small\sl To ``lowest order'' (c.f. our discussion in the next section)
the energy-momentum current ${\cal L}_{\rm tot}^*\vartheta_{\rm
M}\in\Gamma({\rm End}(\tau_{\rm M}))$ of the ``total Lagrangian''}
\bb
\label{vacuumtotallagrangian}
{\cal L}_{\rm tot}(i\ddd - {\rm
M}_{\rm F})(\psi)\equiv {\cal L}_{\rm F}(i\ddd - {\rm M}_{\rm
F})(\psi) + {\cal L}_{\rm D}(i\ddd - {\rm M}_{\rm F})
\ee
{\small\sl reads}
\bb
\label{vacuumenergymomentumtensor}
{\cal L}_{\rm tot}^*\vartheta_{\rm M}\;\sim_{\epsilon\rightarrow 0}\;
\lambda_{\rm gr}\,{\rm tr}{\rm M}^2_{\rm F}\,{\rm id}_{\mbox{\tiny TM}}
+ {\cal O}(\epsilon).
\ee
{\small\sl This holds true for every gauge theory that is based on a
Dirac-Yukawa type operator.}\\

\vspace{0.5cm}

In this section we introduced the Higgs bundle as a specific
Hermitian sub-vector bundle of a chiral fermion bundle and
discussed a specific sub-class of simple type Dirac operators,
called Dirac-Yukawa operators. We also introduced the fermionic
Lagrangian within our geometrical setup. In particular, in the
case of the Lorentzian signature the definition of the fermionic
Lagrangian simply looks like the restriction to the physical
sub-bundle $\xi_{\rm phy}$ of the fermion bundle. However, this is
not the case. In order to obtain the ``correct'' fermionic
couplings one also needs $\xi_{\rm F}^-\subset\xi_{\rm F}.$ Indeed
this doubling of the fermionic degrees of freedom is necessary in
order to consider a Dirac type operator as an endomorphism on the
vector space of sections of a fermion bundle. It is only in this case
that one can make use of the general Lichnerowicz decomposition of
(the square of) a Dirac type operator which in turn permits to consider
the universal Lagrangian (\ref{diraclagrangian}) as a canonical
mapping between the affine set of all Dirac type operators on a
fermion bundle and the top forms of the underlying
base manifold $\mm$.\\

In the next section we will consider a natural generalization of
Dirac-Yukawa type operators which encodes the dynamics of the
sections of the Higgs bundle $\xi_{\rm H}$ and the ``Yang-Mills
bundle'' $\xi_{\rm YM}.$ It also yields the appropriate mass
matrices in such a way that spontaneous symmetry breaking induced
by a minimum of the Higgs potential is in accordance with
spontaneous symmetry
breaking induced by the Yukawa coupling and gravity.

%%%%%%%%%%%%%%%%%%%%%%%%%%%%%%%%%%%%%%%%%%%%%%%%%%%%%%%%%%%%%%%%%%%%%%%%%%%%%%%%

\section[The Lagrangian of PDY-Operators]{The Lagrangian of the
Standard Model as the ``Square'' of Pauli-Dirac-Yukawa Type
Operators}
From our discussion of the preceding section it follows that the
total Lagrangian of a simple type Dirac operator to lowest
order only yields the ``free field'' equations of the eigen sections
of the fermionic mass matrix\footnote{This is because the energy
momentum current is at least homogeneous of degree two with
respect to the appropriate sections.}. Moreover, space-time should
be an Einstein manifold that is physically determined by the (sum
of the) fermionic masses. As a consequence, one has to
appropriately generalize simple type Dirac operators in order to
obtain non-trivial Euler-Lagrange equations also for the
Yang-Mills gauge fields and the sections of the Higgs bundle.
Of course, such a generalization of a simple type Dirac operator
on a fermion bundle must be done in such a way that it is
consistent with spontaneous symmetry breaking induced by the
Yukawa coupling and gravity. For this we first introduce a new
class of Dirac type operators which we call
``Dirac operators of Pauli type'' (PD). These operators act on sections
of a specific sub-bundle of the doubled fermion bundle, where the
latter is defined by the data of a simple type Dirac operator that
underlies the corresponding PD. The doubling of the fermion bundle
has the physical meaning to simultaneously deal with ``particles
and anti-particles''. The above mentioned sub-bundle turns out to
be equivalent to the fermion bundle one starts with and the
corresponding fermionic Lagrangian reduces to the one which is defined
only by the underlying Dirac operator of simple type. To make
this precise, we have to consider real fermion bundles.

\subsection{Real Fermion Bundles and Operators of Pauli Type}
Let $\zeta_{\rm 2F}$ be a real vector bundle of
rank $2\,{\rm N}$ and total space ${\cal W}_{\rm 2F}.$
Also let ${\cal I}_{\rm 2F}\in{\rm End}_\rr(\zeta_{\rm 2F})$
be a
complex structure. We
denote by $\xi_{\rm F}$ the complex vector bundle of rank
${\rm N}$ which is defined by the $\cc-$action:
$z{\goth{z}}:=x{\goth{z}} + y{\cal I}_{\rm 2F}({\goth{z}}),$
for all $z\equiv x+iy\in\cc$ and
$\goth{z}\in{\cal W}_{\rm 2F}.$ The corresponding total space is
denoted again by $\ep.$ Also, let $\xi_{\rm 2F}:=\cc\ot\zeta_{\rm
2F}$ with total space $\ep_{\rm 2F}:=\cc\ot{\cal W}_{\rm 2F}.$ The
complex vector bundle $\xi_{\rm 2F}$ of rank $2\,{\rm N}$
is naturally $\zz_2-$graded since
\bb
\label{complexdecomposition}
\xi_{\rm 2F} \simeq \xi_{\rm F}\op{\overline{\xi_{\rm F}}}.
\ee
Here, ${\overline{\xi_{\rm F}}}$ is the conjugate complex vector
bundle of $\xi_{\rm F}.$ The elements of its total space
${\overline{\ep}}$ are denoted by $\overline{\goth{z}}.$ They may
be identified either with elements $\goth{z}\in{\cal W}_{\rm 2F},$
such that $z{\goth{z}}:=x{\goth{z}} - y{\cal I}_{\rm
2F}({\goth{z}}),$ or considered as anti-linear functionals on
$\ep^*$ (dual of $\ep$). Of course, the subspaces of the
decomposition (\ref{complexdecomposition}) are but the eigen
spaces of ${\cal I}_{\rm 2F}$ (considered as a complex linear
mapping) with respect to the eigenvalues $\pm i.$\\

The canonical real structure on $\xi_{\rm 2F}$ is denoted by
${\cal J}_{\rm 2F}.$ It is given by the action
${\cal J}_{\rm 2F}({\goth{z}}_1,\overline{{\goth{z}}_2}):=
({\goth{z}}_2,\overline{{\goth{z}}_1}).$ The corresponding
real sub-space
\bb
\{(\goth{z},{\overline{\goth{z}}})\in\ep_{\rm 2F}\,|\,\goth{z}\in\ep\}
\simeq{\cal W}_{\rm 2F}
\ee
can be identified with $\ep$ via the canonical complex structure:
$i(\goth{z},{\overline{\goth{z}}}):=(i\goth{z},-i{\overline{\goth{z}}}).$
Note that, likewise, $\ep_{\rm 2F}$ may be viewed as the complex
space
${\cal W}_{\rm 4F}\equiv{\cal W}_{\rm 2F}\op{\cal W}_{\rm 2F}$
with the complex structure given by the action
${\cal I}_{\rm 4F}(\goth{w}_1,\goth{w}_2):=(-\goth{w}_2,\goth{w}_1).$
Clearly, this complex structure in turn can be identified with
${\cal I}_{\rm 2F}$ under the identification of
${\cal W}_{\rm 2F}$ with $\ep.$\\

In what follows, it is assumed that the complex vector bundle
$\xi_{\rm F}$ is a fermion bundle with respect to
$({\rm G},\rho_{\rm F}, D).$ Both, the signature $s\in\zz$ of $D$
and the dimension $n=2k\in\nn$ of the orientable base manifold $\mm$
are again arbitrary, although we are mainly interested in the
physically distinguished case of $(n,s)=(4,\mp2)$. Likewise, the
complex vector bundle $\overline{\xi_{\rm F}}$ is treated as the
conjugate complex (``charge conjugate'') fermion bundle with
respect to $({\rm G},\overline{\rho}_{\rm F},{\bar D}).$ Here,
$\overline{\rho}_{\rm F}$ is the conjugate representation of G and
the (``charge conjugate'') Dirac type operator ${\bar D}$ is
defined by ${\bar D}\overline\psi:= \overline{D\psi},$ for all
$\overline{\psi}\in\Gamma(\overline{\xi_{\rm F}}).$ If
$<\!\cdot,\cdot\!>_\ep$ denotes again the Hermitian
product\footnote{The Hermitian product on $\ep$ is assumed to be
anti-linear in the first, and linear in the second argument. Also,
the ``bar'' notation, as for instance $\overline{\goth{z}},$
should not be confounded with the Dirac conjugation in the case of
the Lorentz signature.} on $\ep,$ then
$<\!\overline{\goth{z}}_1,\overline{\goth{z}}_2\!>_{\overline\ep}
\;:=\;<\!{\goth{z}}_2,{\goth{z}}_1\!>_\ep.$ Hence, the sum
$<\psi,D\psi>_\ep +
<\overline{\psi},{\bar D}\overline\psi>_{\overline\ep}$
vanishes if $D$ is anti-symmetric.\\

Although they are anti-isomorphic to each other, there is no natural
way to identify the fermion bundle $\xi_{\rm F}$ with its charge
conjugate ${\overline{\xi_{\rm F}}}.$ In order to do so we still
have to give additional input. For this let ${\cal J}$ be a
real structure on $\xi_{\rm F}$ such that
\bb
\label{chargeconjugation}
{\cal C}:\,\ep&\longrightarrow&{\overline\ep}\cr
\goth{z}&\mapsto&\overline{{\cal J}(\goth{z})}
\ee
defines a linear bundle isomorphism over the identity on $\mm$, usually
referred to as ``charge conjugation'' (see, for instance, in \cite{BT:88}
in the context of Clifford algebras and in \cite{Con:95} in the context
of non-commutative geometry). Notice that
${\cal C}^{-1}(\overline{\goth{z}})={\cal J}(\goth{z}).$
Then, the charge conjugate Dirac operator may be written as \bb
{\bar D} = {\cal C}_{\rm J}\circ D\circ{\cal C}_{\rm J}^{-1}, \ee
where ${\cal C}_{\rm J}(\goth{z}):={\cal C}({\cal J}(\goth{z}))=
\overline{\goth{z}}.$\\

The existence of ${\cal J}$ depends on the topology of $\xi_{\rm
F}.$ Indeed, it can be shown that a complex vector bundle
possesses a real structure if and only if all of its odd Chern
classes vanish (see, for instance, in \cite{GS:99}).

\begin{definition}
\label{paulidiracop} Let $\xi_{\rm F}$ be a real fermion bundle
over $\mm$ with respect to the Dirac triple
$({\rm G}, \rho_{\rm F},D).$ Also, let
$F_{\!D}\in\Omega^2(\mm,{\rm End}^+_{\rm
Cl}(\ep))$ be the twisting curvature of $\partial_{\!D}.$ We
call the associated first order differential operator
\bb
\label{paulitypedop}
D_{\rm P}:=\left(%
\begin{array}{cc}
  D + i\gamma(F_{\!D}) & 0 \\
  0 & {\cal C}_{\rm J}^{-1}\circ
  (\overline{D - i\gamma(F_{\!D})})\circ{\cal C}_{\rm J} \\
\end{array}%
\right):\,
\begin{array}{c}
  \Gamma(\xi_{\rm F}) \\
  \op \\
  \Gamma(\xi_{\rm F}) \\
\end{array}
\longrightarrow
\begin{array}{c}
  \Gamma(\xi_{\rm F}) \\
  \op \\
  \Gamma(\xi_{\rm F}) \\
\end{array}
\ee
a Dirac operator of ``Pauli type'' (or ``Pauli-Dirac
operator'') with respect to the grading involution
$\Gamma_{\rm 2F}$ that is defined by the action
$\Gamma_{\!\rm 2F}(\goth{z}_1,\goth{z}_2):=
(\Gamma(\goth{z}_2),\Gamma(\goth{z}_1))$ and the real
structure ${\cal J}.$
\end{definition}

Equivalently, one may also express a Pauli-Dirac operator with
respect to the diagonal representation of the grading involution
$\Gamma_{\rm 2F}$ (i.e., where $\Gamma_{\rm 2F}={\rm
diag}(\Gamma,-\Gamma)),$ in which case
\bb
D_{\rm P} &=&
\left(%
\begin{array}{cc}
  D & -\gamma(F_{\!D}) \\
  \gamma(F_{\!D}) & D \\
\end{array}%
\right)\\[0.25cm]\nonumber
&\equiv& D + {\cal I}\ot\gamma(F_{\!D}).
\ee
The bundle mapping
${\cal I}\in{\rm End}_\cc(\ep\op\ep),$ which is defined by
${\cal I}(\goth{z}_1,\goth{z}_2):=(-\goth{z}_2,\goth{z}_1),$
corresponds to the complex structure ${\cal I}_{\rm 4F}$ with
help of the identification
of ${\cal W}_{\rm 2F}\subset\ep\op\overline{\ep}$ with $\ep.$\\

If $D\equiv\ddda,$ then the zero order term $D_{\rm P}-\ddda$
formally looks like the well-known ``Pauli-term''
$i\gamma(F_{\!\rm A})$ which has been introduced by physicists in
order to correctly describe the anomalous magnetic moment of the
proton. However, the first order operator $\ddda +
i\gamma(F_{\!\rm A})$ is not a Dirac type operator in our sense
for the Pauli term is an even operator. To remedy this flaw we
again have to ``double the fermionic degrees of freedom'', in
this case, however, by adding the corresponding
``anti-fermions''. As a consequence, for diagonal sections,
which one may physically interpret as representing the state of a
``particle-anti-particle''\footnote{With help of the identification
$\overline{\xi_{\rm F}}\simeq\xi_{\rm F}$},
$\Psi\equiv(\psi,\psi)\in\Gamma(\xi_{\rm F})\op\Gamma(\xi_{\rm F})=
\Gamma(\xi_{\rm F}\op\xi_{\rm F})\simeq\Gamma(\xi_{\rm 2F}),$
we obtain the identity
\bb
<\Psi,D_{\rm P}\Psi>_{\ep_{\rm 2F}}\; =\; 2<\psi,D\psi>_\ep.
\ee
Hence, the Pauli term does not contribute to the fermionic
Lagrangian as far as ``particle-anti-particle states'' are
simultaneously taken into account. This is certainly desirable for
it is well-known that the coupling of the fermions to the
curvature actually spoils the theory of their renormalizability.
Hence, to lift the first order differential operator $\ddda +
i\gamma(F_{\!\rm A})$ to a ``true'' Dirac type operator restores a
basic feature of (perturbative) quantum field theory. Again, by
formal similarity we also refer to the operator $\ddda +
i\gamma(F_{\!\rm A})$ itself as a Dirac operator of Pauli type,
analogous to the operator (\ref{diracyukawaopplus})
is formally referred to as Dirac operator of Yukawa type.\\

Let $\xi_{\rm F}$ be the real chiral fermion bundle with respect
to $({\rm G},\rho_{\rm F},D),$ with $D$ being of simply type.

\begin{proposition}
\label{ehymhlagrangian} The top form of $D_{\rm P}$ decomposes
into the sum \bb \label{ehymhdecomposition} {\cal L}_{\rm
D}(D_{\rm P}) = {\cal L}_{EH}\pm{\cal L}_{YM}\pm{\cal L}_{H} \ee
where, respectively, ${\cal L}_{EH}$ is the Einstein-Hilbert
Lagrangian, ${\cal L}_{YM}$ the Yang-Mills Lagrangian and ${\cal
L}_{H}$ the ``Higgs'' Lagrangian of the Standard Model of Particle
Physics.
\end{proposition}

\noindent
{\bf Proof:} The proof is basically a copy of the proof
of the corresponding statement that has been presented already in
\cite{Tol:98} in the case of $s=n$ (c.f. Theorem 1). We note
that the top form ${\cal L}_{\rm D}(D')\in\Omega^{\rm n}(\mm)$ is
independent of the connection representing $D'\in{\cal D}(\xi_{\rm
F}).$ Hence, one may choose any representative of the connection
class that corresponds to $D_{\rm P}$ to define the Pauli term
$i\gamma(F_{\! A}).$ The relative signs of
(\ref{ehymhdecomposition}) depend on the signature of $D$ and of
the definition of the Clifford multiplication. In particular, the
relative sign in front of the kinetic term $<\!\partial_{\!
A}^{W}\!\phi,\partial_{\! A}^{W}\!\phi\!>$ of the Higgs Lagrangian
depends on whether $\tau_{\rm Cl}$ or $\tau_{\rm Cl}^{\rm op}$ is
considered to act on $\xi_{\rm F}.$ Finally, we stress that the
decomposition (\ref{ehymhdecomposition}) is actually independent of
the existence of a real structure on $\xi_{\rm F}.$ In particular,
it does not depend on the choice of ${\cal J}.$
\hfill$\Box$\\

The top form (\ref{ehymhdecomposition}) clearly reduces to the
combined Einstein-Hilbert-Yang-Mills Lagrangian in the case where
$\xi_{\rm F}$ is not chiral. However, if $D$ denotes a Dirac-Yukawa
type operator, then
\bb
\label{stmlagrangian}
{\cal L}_{\rm tot}(D_{\rm P})(\Psi) &\equiv&
{\cal L}_{\rm F}(D_{\rm P})(\Psi) +
{\cal L}_{\rm D}(D_{\rm P})\cr &=& {\cal L}_{\rm F}(D)(\psi) +
{\cal L}_{\rm D}(D_{\rm P})
\ee
equals the total Lagrangian of the Standard Model, including
Einstein's theory of gravity. Here, we used the homogeneity property
of the fermionic density: ${\cal L}_{\rm F}(D_{\rm P})(\lambda\Psi)=
\lambda^2{\cal L}_{\rm F}(D_{\rm P})(\Psi)$ and put
$\Psi\equiv(\psi,\psi)/\sqrt{2}.$ Note that
the corresponding Euler-Lagrange equations form a dynamically
closed system. For this reason, we refer to $D_{\rm P}$ also as a
Dirac operator of ``Pauli-Yukawa'' type (or ``Pauli-Dirac-Yukawa''
operator, PDY) if the operator (\ref{paulitypedop}) is defined in
terms of a Dirac-Yukawa type operator (\ref{diracyukawaop}).
Therefore,
\bb
\label{geodataofstm}
(\xi_{\rm F},D_{\rm P})
\ee
may be regarded as a ``square root'' of (the Lagrangian of) the Standard
Model\footnote{Of course, the data $(\xi_{\rm F},D_{\rm P})$ covers
the geometrical properties of the Standard Model only up to the
``semi-classical approximation'' of the latter. It also seems worth
noting that because the decomposition (\ref{ehymhdecomposition}) is
independent of the existence of the reality of the fermion bundle,
it is possible to also take into account magnetic monopoles
within the Standard Model as topologically non-trivial ground states
of the Higgs boson. Moreover, it is well-known that the weak
interaction actually spoils the symmetry under charge
conjugation.}.

\subsection{``Fluctuation'' of a Fermionic Vacuum and the YM-Mass Matrix}
Before we proceed let us come back to the notion of a
``(semi-classical) fermionic vacuum'' and how this is related to
the reality of a fermion bundle. Essentially, a chiral fermion
bundle $\xi_{\rm F}=\xi^+_{\rm F}\op\xi^-_{\rm F}$ is related
to a Dirac triple $({\rm G},\rho_{\rm F},D),$ with $D$ being of
simple type. The existence of a fermionic vacuum crucially depends
on the existence of a non-vanishing section
$\phi\in\Gamma({\rm End}^-_{\rm Cl}(\xi_{\rm F}))$ and a purely
topological Clifford connection $A\in{\cal A}_{\rm Cl}(\xi_{\rm F}).$
This in fact reduces the above Dirac triple to $({\rm H},\rho_{\rm F,red},\ddd)$
and $\xi_{\rm F}$ may be regarded, accordingly, as a perturbation
of the corresponding $\xi_{\rm F,red}.$ Clearly, such a reduction
gives sever topological restrictions on a fermion bundle. Of course,
this holds true also for the existence of a Dirac-Yukawa type
operator. For example, in the case of the electroweak interaction
a fermionic vacuum exists if and only if the corresponding Yang-Mills
gauge bundle of the electroweak interaction is trivial. This in turn holds
true if and only if the (charged) electroweak vector bosons are charge conjugate
to each other (c.f. \cite{Tol:05}). In the (algebraic) torsion free case
this is equivalent to the existence of a flat Yang-Mills connection. This
example may motivate the following

\begin{definition}
A fermion bundle $\xi_{\rm F}$ is called ``perturbative'' provided
there is a Dirac type operator $D\in{\cal D}(\xi_{\rm F})$ such that
${\cal F}_{\!D} = {/\!\!\!\!R}.$
\end{definition}

A fermionic vacuum is thus geometrically described by a
perturbative massive fermion bundle. Next, we introduce a specific
sub-vector bundle of $\xi^*_{\rm F}\ot_\mm\xi_{\rm F}$ and
discuss the ``bosonic mass matrix'' within the presented fermionic frame.

\begin{definition}
\label{yangmillsbdl} Let again $\xi_{\rm F}$ be a massive fermion
bundle with respect to a Dirac-Yukawa model $({\rm G},\rho_{\rm
F},D).$ The real sub-bundle
\bb
\label{ymbdl}
\xi_{\rm YM}:=\tau_{\rm M}^*\ot_\mm{\rm End}^+_{\rm Cl}(\xi_{\rm F})
\subset{\rm End}(\xi_{\rm F})
\ee
is called the ``Yang-Mills bundle'' with respect to the appropriate
fermionic vacuum $\xi_{\rm F,red}.$
\end{definition}

With respect to a fermionic vacuum the (real form of the) Higgs
bundle decomposes into the Whitney sum of two real vector bundles
\bb
\label{higgsdecomposition}
\xi_{\rm H}\simeq\xi_{\rm G}\op\xi_{\rm H,phys}
\ee
with $\xi_{\rm G}\subset\xi_{\rm H}\subset\xi_{\rm F}$ being the
``Goldstone bundle'' and $\xi_{\rm H,phys}\subset\xi_{\rm F}$
being the ``physical Higgs bundle'' (c.f. Lemma 3.1 in \cite{Tol:03a}
for Yang-Mills-Higgs gauge theories). Therefore, any Dirac-Yukawa type
operator on a massive fermion bundle $\xi_{\rm F}$ is
parameterized by $(A,\varphi_{\rm H})\in\Gamma(\xi_{\rm
YM}\times_\mm\xi_{\rm H,phys}).$ In particular, for $t\in[0,1]$ one
may consider the one-parameter family $(A_t,\varphi_t)\in{\cal
A}(\xi_{\rm H})\times\Gamma(\xi_{\rm H})$ which is defined by
$\partial_{\!A,t}:=\partial + tA,$ $\varphi_t:={\cal V} +
t\varphi_{\rm H}.$ Hence, the ``Yang-Mills-Higgs pair''
$(A,\varphi_{\rm H})\in\Gamma(\xi_{\rm YM}\times_\mm\xi_{\rm
H,phys})$ may be physically regarded as a ``fluctuation'' of the
corresponding
fermionic vacuum $\xi_{\rm F,red}.$\\

Like in Yang-Mills-Higgs gauge theories a fluctuation
$(A,\varphi_{\rm H})$ of a fermionic vacuum yields a self-adjoint
section ${\!\rm M}_{\rm H}\in\Gamma({\rm End}(\xi_{\rm H}))\subset
\Gamma({\rm End}_{\rm Cl}(\xi_{\rm F}))$ such that the rank of the
Goldstone bundle equals the dimension of the kernel of the ``Higgs
mass operator'' ${\!\rm M}_{\rm H}.$ Moreover, $\xi_{\rm H,phys}$
decompose into the Whitney sum of eigenbundles of the Higgs mass
matrix. Likewise, since in general $A\in\Gamma(\xi_{\rm YM})$
gives rise to a connection on $\xi_{\rm F}$ that is not compatible
with the fermionic vacuum (i.e. the corresponding covariant
derivative does not commute with the total fermionic mass
operator), a fluctuation of the fermionic vacuum also yields a
non-trivial ``Yang-Mills mass operator'' ${\!\rm M}_{\rm
YM}\in\Gamma({\rm End}(\xi_{\rm YM})).$ (see \cite{Tol:03a}). As a
consequence, the Yang-Mills bundle decomposes into the
eigenbundles of ${\!\rm M}_{\rm YM}$ for again ${\rm spec}({\!\rm
M}_{\rm YM})$ is constant throughout $\mm.$ In particular, one
obtains the decomposition (see, again, \cite{Tol:03a})
\bb
\label{higgsdinner}
\xi_{\rm YM} \simeq
\tau^*_{\rm M}\ot_\mm({\goth{ad}}({\cal Q})\op\xi_{\rm G})
\ee
with ${\goth{ad}}({\cal Q})\equiv{\rm Lie}({\cal H}_{\rm YM})$ being
the ``adjoint bundle'' of the reduced frame bundle
${\cal Q}\stackrel{\iota}{\hookrightarrow}{\cal P},$ which is associated
with the fermionic vacuum $\xi_{\rm F,red}$. Since
${\rm rk}({\!\rm M}_{\rm YM})={\rm rk}(\xi_{\rm G})$ the equivalence
(\ref{higgsdinner}) is a geometrical variant of the famous
``Higgs-Dinner''. It follows that $A\in\Gamma(\xi_{\rm YM})$
decomposes into $A=A_{\rm YM} + A_{\rm G}.$ Hence, the deviation from
$A$ being compatible with the fermionic vacuum can be expressed by
\bb
\label{ymmassoperator}
\partial_{\!\!A}^{{\rm End}(\ep)}{\rm M}_{\rm F}=
{\rm{ad}}(A_{\rm G}){\rm M}_{\rm F}.
\ee
As already mentioned, the non-vanishing of the right hand side
(i.e. of $A_{\rm G}\in\tau_{\rm M}^*\ot_\mm\xi_{\rm G}$) yields a
non-trivial Yang-Mills mass operator ${\!\rm M}_{\rm YM}.$ In fact,
one has
\bb
{\!\rm M}_{\rm YM}(A) = {\rm{ad}}({\rm M}_{\rm F})(A)
\ee
with $\|{\!\rm M}_{\rm YM}(A)\|^2 = {\!\rm M}^2_{\rm YM}(A,A)$
and the symmetric bilinear form
\bb
{\!\rm M}^2_{\rm YM}:\,\Gamma(\xi_{\rm YM}\times_\mm\xi_{\rm YM})
&\longrightarrow&{\cal C}^\infty(\mm)\cr
(A,A')&\mapsto&\mbox{\small$\frac{1}{2}$}\,
{\!\rm M}^2_{\rm YM}(T_a,T_b)\,g_{\rm M}(A^a,A'^b).
\ee
Here, respectively, $A \stackrel{\rm loc}{=}A^a\ot T_a,\;
A' \stackrel{\rm loc}{=}A'^a\ot T_a$ and
\bb
\label{ymmassmatrixsquared}
{\!\rm M}^2_{\rm YM}(T_a,T_b)|_x:=
2\|{\cal G}_{\rm Y}\|^2<\!{\cal V}(x),[T_a,T_b]_+{\cal V}(x)\!>_{\ep}
\ee
is the (squared) ``Yang-Mills mass matrix'', with $[\cdot,\cdot]_+$
being the anti-commutator. Note that we used $\xi_{\rm H}\subset\xi_{\rm F},$
such that a vacuum section ${\cal V}$ can also be regarded as a section
of the fermion bundle. We also extensively used the properties of the
Yukawa mapping (\ref{yukawamapping}). In particular, we used that
${\rm{ad}}({\cal D})A={\cal Y}(A{\cal V})$ where, by abuse of notation,
$A$ refers to two different representations. Also note that the eigenvalues of
(\ref{ymmassmatrixsquared}) are actually independent of $x\in\mm.$ Of course,
the rank of (\ref{ymmassmatrixsquared}) equals the rank of the
Goldstone bundle $\xi_{\rm G}\subset\xi_{\rm F}.$
Accordingly, one may re-write (\ref{ymmassoperator})
for a Clifford connection to be non-compatible with the fermionic
vacuum as
\bb
\|\partial_{\!\!A}^{{\rm End}(\ep)}{\rm M}_{\rm F}\|^2 =
2^n\,{\!\rm M}^2_{\rm YM}(A,A).
\ee
That is, the fermionic mass matrix is covariantly constant with
respect to a Clifford connection on a massive fermion bundle iff
this Clifford connection is in the kernel of the Yang-Mills mass
matrix. The latter, of course, is in one-to-one correspondence
with the residual gauge fields.\\

Let $D\in{\cal D}(\xi_{\rm F})$ be a Dirac operator of simple
type such that $D - \ddda \not= 0$ and ${\cal G}_{\rm YM}$ acts
transitively on the image of $D - \ddda.$ Then, there is a
non-vanishing smooth function $\chi\in{\cal C}^\infty(\mm)$
such that $D = \ddda + i\chi\,{\rm M}_{\rm F}.$ Let
$\xi_{\rm F,red}\simeq\xi_{\rm F}$ be a fermionic vacuum
with respect to $({\rm H},\rho_{\rm F,red},\ddd_{\!\!{\cal D}}).$
Then, $D$ defines a fluctuation of $\xi_{\rm F,red}$ iff
\bb
D = \ddda + \chi(\ddd_{\!\!{\cal D}} - \ddd).
\ee
Note that this condition is in full accordance with the usual
definition of the Higgs boson to be in the ``unitary gauge''.
Here, however, this condition is expressed purely in terms of fermions.

\begin{proposition}
Let $\xi_{\rm F,red}\simeq\xi_{\rm F}$ be a fermionic vacuum with
respect to a Dirac-Yukawa model
$({\rm H},\rho_{\rm F,red},\ddd_{\!\cal V}).$ Also, let
$(A,\varphi_{\rm H})\in\Gamma(\xi_{\rm YM}\times_\mm\xi_{\rm H,phys})$
be a fluctuation of the fermionic vacuum. Then, the total
curvature on $\xi_{\rm F}$ of the connection determined by the
Dirac-Yukawa operator
\bb
D &=& \ddda + \gamma_{\rm M}\ot\phi\cr
&=&
\ddd + \gamma_{\rm M}\ot{\cal Y}({\cal V}) + \gamma(A) +
\gamma_{\rm M}\ot{\cal Y}(\varphi_{\rm H})\cr
&\equiv&
\ddd_{\!\!\cal D} + \gamma(A_{\rm fl})
\ee
reads
\bb
\label{curvaturedecomposition}
{\cal F}_{\!D} &=& \;{/\!\!\!\!R} + F_{\! A} + F_{\rm H} + F_{\!\rm mass}\cr
&=&
\;{/\!\!\!\!R} + F_{\rm YM} + F_{\rm G} + F_{\rm H} + F_{\!\rm mass}.
\ee
Here, respectively,
\bb
F_{\rm YM} &:=& \partial A_{\rm YM} + A_{\rm YM}\wedge A_{\rm YM},\cr
F_{\rm G}  &:=& \partial A_{\rm G} + A_{\rm G}\wedge A_{\rm G},\cr
F_{\rm H}  &:=& \partial A_{\rm H} + A_{\rm H}\wedge A_{\rm H}
\ee
are the Yang-Mills curvature with respect to the reduced Yang-Mills gauge group
${\cal H}_{\rm YM}\subset{\cal G}_{\rm YM}\subset{\cal G}_{\rm F},$
the curvature on $\xi_{\rm F}$ of the (massive) vector boson that corresponds to
the Goldstone boson and the curvature induced by the (physical part of
the) Higgs boson according to the decomposition
\bb
A_{\rm fl} &=& A + A_{\rm H}\cr
&=&
A_{\rm YM} + A_{\rm G} + A_{\rm H},
\ee
with $A_{\rm H}:={\rm ext}_\Theta(\gamma_{\rm M}\ot{\cal Y}(\varphi_{\rm H})).$\\

Finally, the ``mass-curvature'' $F_{\!\rm mass}\in\Omega^2(\mm,{\rm End}(\ep))$
is given by
\bb
F_{\!\rm mass} &:=&
(1-2\|\varphi_{\rm H}\|)\,{\rm M}_{\rm F}^2\Theta\wedge\Theta +
(1+\|\varphi_{\rm H}\|)\,{\rm M}_{\rm YM}(A_{\rm G})\wedge\Theta
\\[0.1cm]\nonumber
&=&
{\rm ext}_\Theta[(1-2\|\varphi_{\rm H}\|)\,\mu_{\rm F} +
(1+\|\varphi_{\rm H}\|)\,\mu_{\rm YM}].
\ee

We call, respectively, $\mu_{\rm F}:=
{\rm ext}_\Theta({\rm M}_{\rm F}^2)\in\Omega^1(\mm,{\rm End}(\ep))$ and
$\mu_{\rm YM}:={\rm M}_{\rm YM}(A_{\rm G})\equiv
\gamma_{\rm M}\ot{\rm M}_{\rm YM}(A)\in\Omega^1(\mm,{\rm End}(\ep))$
the ``fermionic mass form'' and the ``Yang-Mills mass form''.
\end{proposition}

\noindent
{\bf Proof:} First, note that the Yang-Mills mass form $\mu_{\rm YM}$ contributes
to the total curvature even if $F_{\!\rm A}=\varphi_{\rm H}=0$ is supposed
to hold true. Hence, it also gives rise to a ``fluctuation'' of $g_{\rm M}.$
In contrast to what one may infer from $F_{\rm mass}$, however,
the contribution of the bosonic mass is of ``higher order'' in comparison
to the curvature that is induced by the fermionic mass. In other words,
$F_{\!\rm mass} = {\rm ext}_\Theta(\mu_{\rm F}) + {\cal O}(t)$ in accordance
with (\ref{gravindcurvdecomp}). We stress that (\ref{curvaturedecomposition})
indeed reduces to (\ref{gravindcurvdecomp}) if $\varphi_{\rm H}=0.$ Hence,
it gives a physical interpretation, in particular, of the last term of the
decomposition (\ref{gravindcurvdecomp}) of the curvature of a simple type Dirac
operator which spontaneously breaks the gauge symmetry. One may express this
also in more physical terms by saying that it is the interaction of the gauge
field with the fermionic vacuum that yields massive vector bosons.\\

To prove the decomposition (\ref{curvaturedecomposition}) one uses
the decomposition (\ref{higgsdecomposition}) and the Higgs Dinner
(\ref{higgsdinner}), as well as $[{\rm M}_{\rm F}^2,\Theta]=
[{\rm M}_{\rm YM},\Theta]=0.$ Moreover, due to our remark above
concerning fluctuations one may take into account
that $\varphi_{\rm H}=\|\varphi_{\rm H}\|{\cal V}$ (where
$\|{\cal V}\|=1$ is assumed without loss of generality). Note also
that both the (lifted) soldering form $\Theta$ and the Yukawa-mapping
${\cal Y}$ are covariantly constant with respect to any Clifford connection.
Finally, taking also into account that $\partial$ acts on $A_{\rm fl}$
like the usual exterior derivative, the prove actually becomes a
straightforward calculation.\hfill$\Box$\\

\vspace{0.5cm}

We emphasize that spontaneous symmetry breaking induced by
a fermionic vacuum is compatible with spontaneous symmetry
breaking induced by the Higgs potential arising from a
fluctuation of the fermionic vacuum (i.e. $\ddd_{\!\cal D}\mapsto
D_{\rm P}$). Clearly, G acts transitively on ${\rm im}(\phi)
\subset{\rm Orbit}({\goth Z}_0)\simeq{\rm P}\times_{\rm G}{\rm
G}/{\rm H}$ for any chosen minimum ${\goth Z}_0\in{\rm
End}(\cc^{{\rm N}_{\rm H}})$ of the Higgs potential induced by
$D_{\rm P}$. Therefore, the condition $\phi\in\Gamma({\rm
End}(\xi_{\rm H}))\backslash\{\cal O\}$ is necessary and
sufficient for the unitary gauge to exist. In particular, if
$\xi_{\rm F}$ is defined with respect to a Dirac-Yukawa model,
then for each $\varphi\in\Gamma(\xi_{\rm H})\backslash\{\cal O\}$
there exists a ``vacuum section'' ${\cal
V}_{\!\varphi}\in\Gamma({\cal O}rbit({\bf z}_0))\subset
\Gamma(\xi_{\rm H})$ such that $\varphi\in\Gamma(\xi_{\rm
H,phys}).$ This holds true for any rotationally symmetric Higgs
potential (like the Higgs potential generated by a Pauli-Dirac
type operator). By the very definition of the Yukawa mapping
the structure group G then acts transitively also on
${\rm im}({\cal Y}(\varphi)/\|{\cal Y}(\varphi)\|)\subset
{\cal S} \subset{\rm End}_{\rm Cl}(\xi_{\rm F}).$

\section{Outlook}
We discussed a certain class of gauge theories with the basic
property of having a ``square root'' in the sense of the data
of Dirac type operators. These Dirac type gauge theories have in
common that they are derived by a universal Lagrangian which is
shown to be equivariant with respect to bundle automorphisms.
Moreover, these gauge theories naturally include Einstein's theory
of gravity, and the fermionic gauge group of the universal (Dirac-)
Lagrangian contains both Yang-Mills and Einstein-Hilbert type
symmetry groups. In particular, the action of the diffeomorphism
group of the base manifold is naturally represented by pull-back.
We also considered a distinguished class of Dirac type operators
whose associated top form gives rise to spontaneous symmetry
breaking without using Higgs like potentials. Indeed, the latter
naturally arises when a fluctuation of the fermionic vacuum is
taken into account. The geometrical meaning of the induced bosonic
mass operators can be shown to consists of defining the extrinsic
curvature of the ``physical space-time'' $\mm_{\rm phys}.$ The
intrinsic curvature of the latter, however, was shown to be
defined by the fermionic vacuum. In the case where the fermionic
vacuum is defined with respect to a Dirac-Yukawa model, the
appropriate Higgs and Yang-Mills bundle can be naturally regarded
as specific sub-vector bundles of $\xi_{\rm F},$ resp. of
$\xi^*_{\rm F}\ot_\mm\xi_{\rm F}.$ For this
we discussed the Yukawa couplings from a geometrical point of view
in terms of specific sections of the ``Yukawa bundle'' which is
shown to yield the connection between the fermion and the Higgs
bundle. To consider the Yukawa bundle $\xi_{\rm Y}$ as a specific
sub-vector bundle of $\xi^*_{\rm H}\ot_\mm\xi^*_{\rm F}\ot_\mm\xi_{\rm F}$
permits a geometrical understanding of the well-known ``hypercharge relations''
between the physical Higgs boson $\xi_{\rm H,phy}$ and the asymptotically
free fermions $\xi_{{\rm F},{\rm m}^2}\subset\xi_{\rm F}$ in the case of the
minimal Standard Model. In this sense, the presented frame makes it
possible to treat
the geometrical properties of spontaneously broken Yang-Mills-Higgs gauge
theories, as discussed in \cite{Tol:03a}, in terms of fermions. In particular,
it is shown that this kind of gauge theories can be expressed in the
geometrical setup needed to describe fermions without use
of spin structures. Note that the latter actually has no
obvious physical meaning. Indeed, all experiments carried out to date
demonstrating the physical
significance of the two-fold cover of SO(3) are local. The
assumption of orientability, however, is necessary to derive the
Einstein equation from a globally defined density which seems
to also have some significance in our understanding of mass.\\

The ``fermion doubling'' within the presented geometrical setup is
shown to be tied to the Lichnerowicz decomposition of a Dirac type
operator. Since the latter gives rise to the universal Lagrangian
and, moreover, to a specific class of Dirac type operators, which
yield spontaneous symmetry breaking, the projection onto the
physical sub-space $\xi_{\rm phy}\subset\xi_{\rm F}$ clearly
indicates a non-trivial relation between the fermionic Lagrangian
${\cal L}_{\rm F}$ and the Dirac Lagrangian ${\cal L}_{\rm D}.$\\

Since the Dirac Lagrangian is a canonical element within the
presented geometrical frame, it will be useful to discuss it also
in terms of the geometry of variational bi-complexes. This may offer
a more profound mathematical understanding of operators of Pauli-Dirac
type as has been introduced here as a ``fluctuation of a fermionic vacuum''.
These kinds of Dirac type operators obviously play a fundamental
role in the Standard Model of Particle Physics. In a forthcoming
paper we shall thus discuss the Dirac triple of the Standard Model
in more detail. In particular, we shall show how this triple
permits specification of ${\rm spec}({\rm M}_{\rm H}).$
In the case of the ``minimal'' Standard Model ${\rm rk}(\xi_{\rm
H,phys})=1$ which allows a prediction of the mass of the Higgs boson.
For this, however, one still has to carefully take into account possible
``coupling constants'' within the frame of Dirac type gauge theories. In
general, one may modify the total Lagrangian ${\cal L}_{\rm tot}$ as
\bb
{\cal L}_{\rm tot}(D)(\psi)\rightsquigarrow{\cal L}_{\rm phys}(D)(\psi)
:={\cal L}_{\rm F}(D)(\psi) + \lambda\,{\cal L}_{\rm D}(D),
\ee
with the Dirac-Lagrangian being refined by
\bb
{\cal L}_{\rm D}(D) := \ast{\rm tr}(\zeta[D^2-\triangle_{\rm D}]).
\ee
Here, respectively, $\lambda\in\rr$ is a ``relative weight'' between the
fermionic and bosonic Lagrangian and $\zeta$ is the most general
element of the commutant with respect to the fermionic
representation $\rho_{\rm F}$ of the structure group G. More
precisely, $\zeta\in\Gamma({\rm End}_{\rm Cl}^+(\xi_{\rm F}))$ is a
positive Hermitian operator satisfying: $[D,\zeta]=0=[\zeta,g],$
for all $g\in{\cal G}_{\rm YM}.$ It therefore may be considered as
generalizing the Yang-Mills coupling constant of a ``pure''
Yang-Mills gauge theory. Actually, the constant $\lambda$ may be
fixed by an appropriate normalization of the Einstein-Hilbert
Lagrangian.\\

Due to formula (\ref{ymmassmatrixsquared}) the Yang-Mills mass
matrix is proportional to the (squared) norm of the Yukawa-coupling
constants ${\cal G}_{\rm Y}.$ However, the ``physical'' Yang-Mills
mass matrix is known to be proportional to the Yang-Mills coupling
constants $g_{\rm YM}>0$ which parameterize the most general Killing
form on ${\rm Lie(G)}.$ Hence, we have to re-scale $A_{\rm G}$ by
a positive constant $g_{\rm G}$  for each simple factor of ${\rm G},$
i.e. $A_{\rm G}^a\rightsquigarrow A^a_{\rm G}/g^{(a)}_{\rm G}$
(no summation involved), such that
\bb
g^{(a)}_{\rm YM} = g^{(a)}_{\rm G}g_{\rm Y}
\ee
with the abbreviation $g_{\rm Y}\equiv\|{\cal G}_{\rm Y}\|.$\\

Finally, one also has to take into account that in general
$\|{\cal V}\|\not= 1,$ and that the various differential forms defining
the Dirac type operator in question have different dimensions.
Besides the ``Planck scale'' (which comes in because of the generic
Einstein-Hilbert part of the total Lagrangian) this will bring in
an additional length scale within Dirac type gauge theories.
However, in the case of the Lagrangian of a PDY this additional
length scale turns out to be proportional to the (inverse of the)
Higgs mass. Hence, in the case of the Standard Model the two length
scales decouple within GTDT and gravity effects can be neglected as
it is commonly expected. For this to be consistent, however, we
stress again the necessity of the compatibility of the two
different symmetry reductions obtained by the fermionic vacua
(i.e. simple type Dirac operators) and the ground states of the
Higgs boson (i.e. Pauli type Dirac operators).\\

We finish with some (rather) speculative remarks on how ``quantum
corrections'' might be incorporated in Dirac type gauge theories.
For this let again $\xi_{\rm F,red}$ be a fermionic vacuum with
respect to a given Dirac-Yukawa model $({\rm H},\rho_{\rm
F,red},\ddd_{\!\cal V}).$ Accordingly, let $\mm_{\rm
phys}\equiv{\rm im}(\cal V)\subset{\rm E}_{\rm H}\subset\ep$ be
the ``physical space-time'' with respect to the fermionic vacuum.
As mentioned above, the geometry of $\mm_{\rm phys}$ is determined
by ${\rm M}_{\rm F},$ ${\rm M}_{\rm YM}$  and ${\rm M}_{\rm H},$
respectively, in the sense that $g_{\rm M}$
is determined by the spectrum of the fermionic mass operator and
the Higgs potential evaluated with respect to $\cal V.$ The normal
sections of $\mm_{\rm phys}$ are determined by the eigenbundles of
the bosonic mass operators that correspond to the massive bosons.
Hence, a change of the fermionic vacuum leads to a change of the
geometry of $\mm_{\rm phys},$ provided the respective spectra of
the corresponding mass operators are changed. Naively, this will
be caused by ``quantum corrections'' to the propagators of the
``asymptotically free particles'' like, for instance, of
$\ddd_{\!\cal V}^{-1}$ in the case of asymptotically free fermions
(\ref{asymptoticfreefermions}). In this respect, the geometrical
frame presented so far mimics perturbation theory to lowest order
in the Planck-constant $\hslash$. Of course, the task then
consists in expressing ``quantum corrections'' in terms of an
appropriate ``quantum stochastic'' on the moduli space of simple
type Dirac operators of the form (which, however, is known to be
not well-defined for arbitrary signature\footnote{In the case
where $\mm$ is compact and $\ddd_{\!\cal V}$ is elliptic, the
propagator $\ddd_{\!\cal V}^{-1}$ is well-defined in terms of
Fourier integral operators and one may choose, for instance, the
``zeta-function'' to regularize formal expressions like
$\log\det(1+\ddd_{\!\cal V}^{-1}\circ{/\!\!\!\!A}_{\rm fl})$.}):
\bb {\goth W}(g_{\rm M},A,\varphi_{\rm H})\equiv
\log\!\frac{\det_\Lambda\!(\ddd_{\!\cal V} + {/\!\!\!\!A}_{\rm
fl})}{\det_\Lambda\!\ddd_{\!\cal V}}. \ee Here, respectively,
$\Lambda$ is some ``regularizing cut-off'' and $ {/\!\!\!\!A}_{\rm
fl}:=\gamma(A)+\gamma_{\rm M}\ot{\cal Y}(\varphi_{\rm H})\in
\Gamma({\rm End}(\xi_{\rm F})),$ with $(A,\varphi_{\rm
H})\in\Gamma(\xi_{\rm YM}\times_\mm\xi_{\rm H,phys})$ being a
fluctuation of $\xi_{\rm F,red}.$ Notice again, that a quantum
fluctuation of the fermionic vacuum would yield a fluctuation of
both the inner as well as the exterior geometry of $\mm_{\rm
phys}$ and hence a fluctuation
of all bosons. This again emphasizes the geometrical role of fermions.\\

\section{Appendix}
Because of its relevance within Dirac type gauge theories we
present here in some detail the proof of Proposition
(\ref{propsimpletypedop}). In particular, it is shown that it
holds true for arbitrary signature of $D.$ In \cite{AT:96} a
similar proof was presented for the special case of elliptic Dirac
type operators.

\subsection{Tensor Decomposition}
 In this sub-section we collect some useful formulae which will be
 needed to prove the explicit form of the Dirac forms
 $\varpi_{\rm D}\in\Omega^1(\mm,{\rm End}(\ep))$ of simple type
 Dirac operators. Though interesting in its own we will not prove
 these formulae here (since the proof would be technical but
 straightforward).\\

 To get started let $\omega\in\Gamma(\tau_{\rm M}^*\ot_\mm
 \Lambda^{\rm n-1}\tau_{\rm M}^*\ot_\mm{\rm End}(\xi_{\rm F})).$
Throughout this Appendix, let $(X_1,\ldots,X_n)$ be a locally
defined orthonormal frame on $\mm$ and $(X^1,\ldots,X^n)$ its dual
frame. Then, locally one has
\bb
 \omega(X_{i_1},\ldots,X_{{i_n}})
 &=:&
 \om_{i_1\cdots i_n}\equiv\om_{i_1 [i_2\cdots i_n ]},\cr
\gam (\om )
 &=:&
{/\!\!\!\omega}\equiv
 \gam^{i_1}\dots\gam^{i_n}\circ\om_{i_1\cdots i_n}.
\ee
Here, respectively, the brackets $[\cdots]$ indicate
skew-symmetrization with the convention:
$n!\,\om_{[i_1\cdots i_n]}=\sum_{\sig\in S_n}\sgn \sig\,
\om_{\sig(i_1)\cdots\sig(i_n)}$ and, again, $\gamma^k\equiv\gamma(X^k).$
In what follows, we restrict ourselves to the Clifford relation
$\alpha\beta + \beta\alpha = +2g_{\rm M}(\alpha,\beta)$ for all
$\alpha,\beta\in T^*\!M\hookrightarrow Cl(M)$ (the total
space of $\tau_{\rm Cl}$).\\

First, we have the following decomposition\footnote{The "$\uparrow
\atop j$" means that $i_1$ is at the jth position.}
\bb
\label{form1}
   \om_{i_1\cdots i_n} = \om_{[i_1 \cdots i_n]} +\frac{1}{n}\sum^n_{j=2}
                         \left(\om_{i_1\cdots i_n}
                        +\om_{\begin{array}{cccccc}
                               i_j & i_2 &  \cdots & i_1 & \cdots & i_n \\
                                   &     &     & \uparrow   &  &     \\
                                   &     &        &  j         &  &
                              \end{array}}
                          \right) .
 \ee

As a consequence, it follows that $\gamma(\omega)$ may locally be
written as
\bb
 \label{form2}
  \gam^{i_1}\gam^{i_3}\cdots\gam^{i_n}\circ\om_{i_1 \mu i_3\cdots i_n}
  &=&-\,
  \frac{n}{n-1}\,\gam^{i_1}\gam^{i_3}\cdots\gam^{i_n}\circ
  \om_{[\mu i_1 i_3\cdots i_n]}\nonumber\\
   &&+\,
   \frac{1}{n-1}\,\gam^{i_1}\gam^{i_3}\cdots\gam^{i_n}\circ
   \om_{\mu i_1 i_3\cdots i_n}\nonumber\\
   &&-\,
   (n-2)\,g^{\alp\bet}\,\gam^{i_4}\cdots\gam^{i_n}\circ
   \om_{\alp\bet \mu i_4\cdots i_n}
\ee
where, again, $g^{ij}\equiv g_{\rm M}(X^i,X^j).$\\

Using these two formulae one finally proves the following local
decomposition which turns out to be  particularly useful in what
follows: \bb \label{form4}
  \gamma(\omega)
  &\stackrel{loc.}{=}&
  \gam^{i_1}\cdots\gam^{i_n}\circ\om_{[i_1\cdots i_n]}+\,
   (n-1)\,g^{\alp\bet}\,
   \gam^{i_3}\cdots\gam^{i_n}\circ\om_{\alp\bet i_3\cdots i_n}.
\ee

\subsection{Proof of Proposition \ref{propsimpletypedop}}
Let $\xi=(\ep,\mm,\pi_\ep)$ be an arbitrary $\zz_2-$graded
Clifford module bundle over any smooth (semi-)Riemannian manifold
$(\mm,g_{\rm M})$ with $\dim\mm =n$ and $n$ even. Every Dirac type
operator $D$ may be globally decomposed as $D=\ddda +
{/\!\!\!\omega}$ with $A$ being a Clifford connection and
$\omega\in\Omega^1(\mm,{\rm End}^+(\ep))$ being given by
$\omega:=\Theta\wedge(D - \ddda).$ Notice again that
$\omega$ may also depend on the choice of $A$ unless $D$ is of
simple type. Locally, $\omega$ reads
\bb
\omega &\stackrel{loc.}{=}&
X^i\ot\omega^a_i\ot{\goth e}_a\nonumber\\[0,2cm]
&\equiv& X^i\ot\left(\,\sum_{k=0}^n\gamma^{i_1}\cdots\gamma^{i_k}
\omega_{ii_1\cdots i_k}^a\right)\ot{\goth e}_a,
\ee
with $\omega_{ii_1\cdots i_k}^a=\omega_{i[i_1\cdots i_k]}^a$ and
$({\goth e}_1,\ldots,{\goth e}_{\rm N})$ being a local frame in
${\rm End}_{\rm Cl}(\xi)$ such that $\omega$ is odd with respect
to the total grading.\\

By definition, $D$ is of simple type if the Clifford connection
$A$ also defines the Bochner-Laplacian of $D.$ Using the general
Bochner-Lichnerowicz-Weizenb\"ock decomposition of $D^2$ it can be
shown that, independently of the signature of $g_{\rm M},$ this
holds true if and only if\footnote{In the case $s=n$ this
has been proved in \cite{AT:96}. The more general case of
arbitrary signature has been proved in \cite{Thum:02}.}
\bb
 \label{simcon}
2g^{ij}\omega^a_j +\gamma^j [\omega^a_j ,\gamma^i ] =0.
\ee

Since this relation is linear with respect to the frame $({\goth
e}_1,\ldots,{\goth e}_{\rm N})$ we may suppress the index $a$ in
what follows.

\begin{lemma}
\label{sol}
  Let $\omega\in\Gamma(\tau^*_{\rm M}\ot_\mm\tau_{\rm Cl})$ be a
  Clifford algebra valued one-form where the coefficients $\omega_\nu$
  fulfill the relation (\ref{simcon}). Then, the most general form
  of $\omega_\nu$ reads
\bb
   \om_{\nu} = \sum_{k=0}^n \gam^{i_1}\cdots\gam^{i_k}
    \om_{\nu[i_1\cdots i_k]}^{(k)}
 \ee
  where the coefficients satisfy the relations:
  \begin{eqnarray}
   &&\om^{(n)}_{[\nu i_1\cdots i_n ]} = 0,\nonumber \\[0.2cm]
   &&\om^{(n-1)}_{[\nu i_1\cdots i_{n-1}]} =
   \eps_{\nu i_1\cdots i_{n-1}}f,\nonumber \\[0.2cm]
   &&k g^{\alp\bet}\om^{(k)}_{\alp\bet i_1\cdots i_{k-1}}+\om^{(k-2)}_{[i_1\cdots
   i_{k-1}]}=0,\quad k= n-1, \dots ,2,\nonumber \\[0.2cm]
   &&g^{\alp\bet}\om^{(1)}_{\alp\bet} = 0.
  \end{eqnarray}
 Here, respectively,
  $\om^{(k)}_{\nu i_1 \cdots i_k}\equiv\om^{(k)}_{\nu [i_1\cdots i_k ]}:=
  \omega^{(k)}_\nu(X_{i_1},\ldots X_{i_k})$ are the local
  coefficients of appropriate k-forms $\omega^{(k)}_\nu\in\Omega^1(U)$
  ($U\subset\mm$ open, $\nu=1,\ldots ,n$), $f\in{\cal C}^\infty(U)$ and
  $\epsilon_{i_1 \dots i_n}\equiv\mu_{\rm M}(X_{i_1},\ldots,X_{i_n})$
  the Levi-Civita symbol.
 \end{lemma}

\noindent {\bf Proof:} To get started we re-write condition
(\ref{simcon}) as $\gamma^\mu\gamma^\nu\omega_\nu +
\gamma^\nu\omega_\nu\gamma^\mu = 0$ and then appropriately
re-arrange both terms on the left hand side. \bb
  \gamn\om_{\nu}\gamm =
  \sum_{k=0}^n (-1)^k (\gamma^\nu\gamma^\nu
  \gam^{i_1}\cdots\gam^{i_k}\omega^{(k)}_{\nu i_1\cdots i_k}
  -2k\,g^{\mu i_1}\gamn\gam^{i_2}\cdots\gam^{i_k}
  \om^{(k)}_{\nu i_1 i_2\cdots i_k}).
 \ee

 Using this re-arrangement and formula (\ref{form2}) one obtains:
 \bb
  0 &=&
  \sum_{k=0}^n \left( (1-(-1)^k )
  (\gamm\gamn\gamma^{i_1}\cdots\gamma^{i_k}\omega^{(k)}_{[\nu i_1\cdots i_k]}
  + k\gamm\gam^{i_2}\cdots\gam^{i_k}g^{\alp\bet}
  \omega^{(k)}_{\alp\bet i_2\cdots i_k}\,)\right.\nonumber\\[0.2cm]
  &&\qquad
  +(-1)^k\,2(k+1)\,g^{\mu\nu}
  \gamma^{i_1}\cdots\gamma^{i_k}\omega^{(k)}_{[\nu i_1\cdots i_k]}\nonumber\\[0.2cm]
  &&\qquad
  +\left.(-1)^k\,2k(k-1)\,g^{\mu\nu}\gam^{i_3}\cdots\gam^{i_k}
  g^{\alp\bet}\omega^{(k)}_{\alp\bet\nu i_3\cdots i_k}\,\right).
\ee

This sum may be further split into two sums of an even and odd
number of Clifford elements. Since these terms are linearly independent
one may evaluate each sum separately. For example, the sum of an
odd number of Clifford elements gives rise to the condition:
\bb
\label{oddsum}
0\!\!&=& \!\!\sum_{\stackrel{k=1}{\mbox{\tiny(k odd)}}}^n
\left(\gamm\gamn\gamma^{i_1}\cdots\gamma^{i_k} \omega^{(k)}_{[\nu
i_1\cdots i_k]}\right. +k\,\gamm\gam^{i_2}\cdots\gam^{i_k}
\gab\omega^{(k)}_{\alp\bet i_2\cdots i_k}\nonumber\\[0.2cm]
&&\quad -(k+1)\,g^{\mu\nu}\gamma^{i_1}\cdots\gamma^{i_k}
\omega^{(k)}_{[\nu i_2\cdots i_k]}\left.
-k(k-1)\,g^{\mu\nu}\gam^{i_3}\cdots\gam^{i_k}
\gab\omega^{(k)}_{\alp\bet\nu i_3\cdots i_k}\right).
\ee

Since\footnote{The $\hat{\cdot}$ denotes the omission of the "hated"
object.} $\gamm\gamn\gam^{i_1}\cdots\gam^{i_{n-1}}
\om^{(n-1)}_{[\nu i_1\cdots i_{n-1}]} =
n\,g^{\mu\nu}\gam^{i_1}\cdots\hat{\gam^{i_{\mu}}}\cdots\gam^{i_n}
\om^{(n-1)}_{[\nu i_1 \cdots \hat{i_{\mu}}\cdots i_n]}$,
the condition (\ref{oddsum}) becomes equivalent to
\bb
0&=& \gamm\gam^{i_2}\cdots\gam^{i_{n-1}}
\left((n-1)\gab\om^{(n-1)}_{\alp\bet i_2\cdots i_{n-1}}+
\om^{(n-3)}_{[i_2\cdots i_{n-1}]}\,\right)\nonumber\\[0.2cm]
&& +\sum_{\stackrel{k=3}{\mbox{\tiny(k odd)}}}^{n-3} \left(
-(k+1)\,g^{\mu\nu}\gam^{i_1}\cdots\gam^{i_k}((k+2)\gab
\om^{(k+2)}_{\alp\bet\nu i_1 \cdots i_k}+
\om^{(k)}_{[\nu i_1 \cdots i_k]})\right.\nonumber\\[0.2cm]
&&\left.\qquad\quad
+\,\gamm\gam^{i_2}\cdots\gam^{i_k}(k\gab\omega^{(k)}_{\alp\bet
i_2\cdots i_k}+ \om^{(k-2)}_{[i_2\cdots i_k]})\right) +\,
\gamm\gab\om^{(1)}_{\alp\bet}.
\ee

The term with the highest degree in the $\gam^{i_j}$ vanishes. By
an induction argument one ends up with the recursion
relation:
\bb
k\,\gab\omega^{(k)}_{\alp\bet i_2\cdots i_k}+
\om^{(k-2)}_{[i_2\cdots i_k]}=0,\quad k=3,\dots n-1. \ee As a
consequence, it follows that $\gab \om_{\alp\bet}^{(1)}=0.$
Moreover, the term $\om^{(n-1)}_{[\nu i_1\dots i_{n-1}]}$ drops
out and thus is undetermined. Its most general form is given by
\bb \om^{(n-1)}_{[\nu i_1\dots i_{n-1}]} = \eps_{\nu i_1\dots
i_{n-1}} f,
\ee
with $f$ being an arbitrary locally defined smooth function on $\mm.$\\

Next, we consider the sum of an even number of Clifford elements.
This yields the relation
\bb
\sum_{\stackrel{k=0}{\mbox{\tiny(k even)}}}^{n}
\left( (k+1)\,g^{\mu\nu}\gamma^{i_1}\cdots\gamma^{i_k}
\omega^{(k)}_{[\nu i_1\cdots i_k]} +\,
k(k-1)\,g^{\mu\nu}\gam^{i_3}\cdots\gam^{i_k}\gab
\omega^{(k)}_{\alp\bet\nu i_3\cdots i_k}\right) = 0,
\ee
which in turn gives rise to the following constraint equations:
\bb
0 &=& (n+1)\,g^{\mu\nu}\gam^{i_1}\cdots\gam^{i_{n}}
\om^{(n-1)}_{[\nu i_1\cdots i_{n}]},\nonumber\\[0.2cm]
0 &=& (n-1)\,g^{\mu\nu}\gam^{i_1}\cdots\gam^{i_{n-2}}
(n\gab\om^{(n)}_{\alp\bet\nu i_1\cdots i_{n-2}}+
\om^{(n-2)}_{[\nu i_1\cdots i_{n-2}]}),\nonumber\\[0.2cm]
&\vdots&\nonumber\\[0.2cm]
0 &=& (k+1)\,g^{\mu\nu}\gamma^{i_1}\cdots\gamma^{i_k}
((k+2)\gab\om^{(k+2)}_{\alpha\beta\nu i_1\cdots i_k}+
\omega^{(k)}_{[\nu i_1\cdots i_k]}),\nonumber\\[0.2cm]
&\vdots&\nonumber\\[0.2cm]
0 &=&
g^{\mu\nu}\gam^i\gam^j(2\gab\om^{(2)}_{\alp\bet\nu}+\om^{(0)}_{\nu}).
\ee

These are satisfied provided that \bb 0 &=&
\om^{(n)}_{[\nu i_1\cdots i_{n}]}\nonumber\\[0.2cm]
0 &=& (k+2)\gab\om^{(k+2)}_{\alp\bet\nu i_1\cdots i_k} +
\omega^{(k)}_{[\nu i_1\cdots i_k ]},\quad k=0,\dots ,n \ee which,
when combined with our previous result with respect to the sum of
an odd number of Clifford elements, finally proves the
statement. \hfill$\Box$\\

\begin{korollar}
Let $\xi_{\rm F}$ be the chiral fermion bundle with respect to the
Dirac triple $({\rm G},\rho_{\rm F},D),$ with $D$ being of simple
type and of arbitrary signature. The Dirac form of $D$ reads
$\varpi_{\rm D}=\Theta\wedge(\gamma_{\rm M}\ot\phi),$ with
$\phi\in\Gamma({\rm End}_{\rm Cl}^-(\xi_{\rm F}))$ uniquely
determined by $D.$
\end{korollar}

\noindent {\bf Proof:} Again, in the sequel we shall suppose that the
induced Clifford relations, defining $\tau_{\rm Cl},$ are given by
$\alpha\beta + \beta\alpha = +2g_{\rm M}(\alpha,\beta).$ Locally,
we may write $\varpi_{\rm D}(X_\mu)=\omega^a_\mu\ot{\goth e}_a$
and, again, decompose the coefficients into the sum of odd and
even terms with respect to the canonical involution $\alpha\mapsto
-\alpha$ for all $\alpha\in T^*\!M\hookrightarrow Cl(M)$:
 \bb
  \omega^a_\mu
   &=& \sum^n_{k=1} \gamma^{i_1}\cdots\gamma^{i_k}
       \omega^a_{\mu [i_1\cdots i_k]}\nonumber\\[0.2cm]
   &=&
       \sum^{n-1}_{\stackrel{k=1}{\mbox{\tiny (k odd)}}}
       \gamma^{i_1}\cdots\gamma^{i_k}\omega^a_{\mu [i_1\cdots i_k]}
      +\sum^n_{\stackrel{k=0}{\mbox{\tiny (k even)}}}
      \gamma^{i_1}\cdots\gamma^{i_k}\omega^a_{\mu [ i_1\cdots i_k]}\nonumber\\[0.2cm]
   &\equiv&
      \alpha^a_{\mu} + \beta^a_{\mu} .
 \ee
 We then compute $\gamma^\mu\omega_\mu^a\equiv\gamma^\mu\alpha^a_\mu +
 \gamma^\mu\beta^a_\mu$ to show that $\gamma^\mu\omega^a_\mu\ot{\goth e}_a=
 \gamma_{\rm M}\ot\phi.$\\

 With help of formula (\ref{form4}) one obtains\footnote{For notational
 convenience the index $a$ is again suppressed.}
 \bb
  \gamma^\mu\alpha_\mu
  &=&
  \gamma^\mu\gamma^{i_1}\cdots\gamma^{i_{n-1}}
  \omega^{(n-1)}_{[\mu i_1\cdots i_{n-1}]} + 2\,g^{ij}\omega^{(1)}_{ij}\nonumber\\[0.2cm]
  &+&
  \sum^{n-3}_{\stackrel{k=1}{\mbox{\tiny(k odd)}}}
  \left((k+2)\,\gamma^{i_2}\cdots\gamma^{i_{k+2}}
  g^{ij}\omega^{(k+2)}_{ij[i_2\cdots i_{k+2}]}
  + \gamma^\mu\gamma^{i_1}\cdots\gamma^{i_k}
 \omega^{(k)}_{[\mu i_1\cdots i_k]}\,\right).
 \ee
Hence, using Lemma \ref{sol}, one concludes that
 \bb
 \gamma^\mu\alpha_{\mu} = \gamma^\mu\gamma^{i_1}\cdots\gamma^{i_{n-1}}
\omega^{(n-1)}_{[\mu i_1\cdots i_{n-1} ]}.
 \ee

Next, we consider $\gamma^\mu\beta_\mu$ and find, using similar
arguments like above, that
 \bb
\gamma^\mu\beta_\mu &=& \gamma^\mu\gamma^{i_1}\cdots\gamma^{i_n}
\omega^{(n)}_{[\mu i_1\cdots i_n]}\nonumber\\[0.2cm]
&+& \sum^{n-2}_{\stackrel{k=0}{\mbox{\tiny(k even)}}}
\left((k+2)\,\gamma^{i_2}\cdots\gamma^{i_{k+2}}\,g^{ij}
\omega^{(k+2)}_{ij[i_2\cdots i_{k+2}]} +
\gamma^\mu\gamma^{i_1}\cdots\gamma^{i_k}
\omega^{(k)}_{[\mu i_1\cdots i_k]}\,\right)\nonumber\\[0.2cm]
&=& 0.
 \ee

Finally, using Lemma \ref{sol} again, we end up with
 \bb
 \gamma^\mu\omega_\mu
  &=&
  \gamma^\mu\gamma^{i_1}\cdots\gamma^{i_{n-1}}
  \omega^{(n-1)}_{[\mu i_1\cdots i_{n-1}]}\nonumber\\[0.2cm]
  &=&
f\,\gamma^1\cdots\gam^n\nonumber\\[0.2cm]
&=& {\tilde f}\,\gamma_{\rm M}.
 \ee

If we set $\phi\equiv {\tilde f}^a{\goth e}_a,$ where
$({\goth e}_1,\ldots{\goth e}_{{\rm N}^-})$ is a local frame in
${\rm End}^-_{\rm Cl}(\xi_{\rm F}),$ we obtain the desired result and
thus have also proved Proposition \ref{propsimpletypedop} in the
case of $\tau_{\rm M}$. Of course, for $\tau_{\rm Cl}^{\rm op}$
the proof is similar.\hfill$\Box$

\vspace{1.85cm}

\noindent
{\bf Acknowledgments}\\
The authors would like to thank E. Binz for very interesting
and stimulating discussions on the presented subject. Also,
one of the authors (JT) would like to thank K. Marathe for his
interests and valuable comments.

\end{document}